\documentclass[review]{elsarticle}

\usepackage{lineno,hyperref}
\modulolinenumbers[5]

\usepackage{graphicx}
\usepackage{dcolumn}
\usepackage{bm}
\usepackage{multirow}
\usepackage{textcomp}
\usepackage[]{subfigure}

\usepackage{booktabs}
\usepackage{array}
\usepackage{pgf}
\usepackage{graphicx}
\usepackage[absolute,overlay]{textpos}
\usepackage{calc}
\usepackage{amsmath}
\usepackage{amsfonts}
\usepackage{amssymb}
\usepackage{textcomp}
\usepackage{tikz}
\usepackage{gensymb}
\usepackage{color,soul}
\usepackage{xcolor}

\newcommand{\commentCWS}[1]%
{\textsf{\textcolor{blue}{#1$^{\mathrm{CWS}}$}}}
\usepackage[colorinlistoftodos]{todonotes} 

\begin{document}

\begin{frontmatter}

\title{A Scale-Separated Full-Field Monopole Method for Precipitation Modelling}

\author[ubc]{C.~W.~Sinclair} 

\address[ubc]{Department of Materials Engineering, The University of British Columbia, 309-6350 Stores Road, Vancouver, Canada} 

\begin{abstract}
A scale-separated monopole method is developed for full-field prediction of precipitation under conditions where particle-scale diffusion fields coexist with solute variations over much larger distances. Precipitates are represented by monopole Green's-function solutions, providing a particle-based description of interacting particle growth and dissolution. The direct quasistatic formulation is validated against transient finite-difference calculations and a conventional Kampmann--Wagner numerical model, but becomes inappropriate when spatially heterogeneous particle populations generate slowly relaxing long-wavelength concentration fields. An Ewald decomposition is therefore used to separate the diffusion field by spatial scale. Screened particle-scale fields are treated quasistatically, while the complementary long-range field evolves by transient diffusion. The resulting hybrid formulation retains explicit particle-level precipitation kinetics while allowing mesoscale solute redistribution to evolve on its physical timescale. Application to Al--Sc demonstrates that a transient depletion field generated by localized precipitation can suppress subsequent nucleation and produce persistent spatial variations in precipitate number density and volume fraction.
\end{abstract}


\end{frontmatter}

\date{\today}


\section{Introduction}

Predictive precipitation models range from computationally efficient mean-field models \cite{Robson2004ModellingPrecipitation,Perez2008ImplementationPrecipitation,Deschamps1999InfluenceStress} to spatially resolved full-field simulations \cite{Eymann2024AModel,Fallah2012Phase-fieldAlloys,Kim2017First-principles/Phase-fieldAlloys}. Mean-field approaches are particularly well suited to alloy and heat-treatment modelling because they describe all stages of precipitation at modest computational cost. Modern implementations commonly track the full particle size distribution, as in Kampmann--Wagner numerical (KWN) models \cite{Wagner2007HomogeneousPrecipitation,Robson2004ModellingPrecipitation,Robson2004AScandium,Perez2008ImplementationPrecipitation,Du2012AAlloys}, allowing the evolution of particles at different stages of nucleation, growth, and coarsening to be treated within a unified approach. Their computational efficiency arises from the fact that they do not track the spatial distribution of individual precipitates and the associated local variations in matrix composition.

A central ingredient of such mean-field models is the growth law for an individual precipitate, commonly based on the Zener solution for diffusion-controlled growth of a spherical particle \cite{Zener1949TheorySolution}. In this description, the concentration field surrounding an isolated precipitate is assumed to be quasi-steady, and the growth rate is determined by the difference between the interfacial and far-field matrix compositions. Each particle therefore responds to a common matrix composition rather than to the local diffusion fields generated by its neighbours, so diffusional interactions associated with overlapping concentration fields (soft impingement) are typically ignored. More generally, such models cannot directly describe precipitation within a spatially heterogeneous and evolving solute field, as may arise near defects \cite{Deschamps1999InfluenceStress}, from spatial variations in nucleation behaviour \cite{Eymann2024AModel}, or from solute redistribution during earlier stages of a thermal history \cite{Du2013MicrostructureResults}.

Full-field diffuse-interface (phase-field) models represent the evolving microstructure through continuous fields \cite{Fallah2012Phase-fieldAlloys,Kim2017First-principles/Phase-fieldAlloys}, while full-field sharp-interface approaches explicitly track precipitate interfaces and solve for solute transport in the surrounding matrix \cite{Eymann2024AModel}. Both approaches naturally describe spatially varying solute fields and interactions between neighbouring precipitates, but generally require spatial resolution comparable to the scale over which the concentration varies around individual particles. This creates a demanding separation of length scales in dilute alloys, where nanometre-scale precipitates may coexist with interparticle spacings and microstructural concentration variations extending over hundreds or thousands of nanometres. Simulation volumes must be sufficiently large to contain statistically meaningful precipitate populations at realistic number densities.

An alternative is to retain precipitates as discrete particles while representing their diffusion fields analytically using Green's functions. In the lowest-order, or \emph{monopole}, approximation \cite{Voorhees1984,Voorhees1984SolutionSimulations}, the absorption or release of solute by each precipitate is represented by a point sink or source located at its centre, with the source strength directly related to the particle growth rate. The concentration field is then obtained by superposition of the contributions from all particles. Diffusional interactions between spatially distributed precipitates can therefore be retained without explicitly resolving either the particle interfaces or their surrounding concentration fields on a computational mesh. This approach was applied to Ostwald ripening of interacting precipitate populations \cite{Voorhees1984,Voorhees1984SolutionSimulations}. The treatment was generalized through a multipole expansion, establishing the accuracy of the monopole approximation and the increasing importance of higher-order contributions at larger precipitate volume fractions \cite{Akaiwa1994Late-stageFunctions}. Sagui and co-workers later coupled this with homogeneous nucleation, extending the approach to cover nucleation, growth, and coarsening \cite{Sagui1997NucleationState,Sagui1999TheorySeparation}.

Related particle-based Green's-function approaches have subsequently been extended to spatially heterogeneous coarsening problems. Yang et al.~\cite{Yang2017ModelingZones} combined a point-source treatment of diffusional interactions with a spatially varying background concentration to describe precipitate-free-zone formation near grain boundaries. In their formulation, the grain boundary acts as a solute sink and the background concentration varies with distance from the boundary over a time-dependent depletion length. This demonstrated that particle-based diffusion models can capture the coupling between coarsening and mesoscale solute depletion, including the development and broadening of precipitate-free zones.

More generally, however, existing quasistatic monopole formulations do not independently evolve an arbitrary long-wavelength concentration field according to the transient diffusion equation. In the formulation of Sagui and Grant \cite{Sagui1999TheorySeparation}, for example, diffusional interactions between individual particles are retained explicitly, but the background supersaturation is represented by a spatially uniform, time-dependent mean-field quantity. Yang et al.~\cite{Yang2017ModelingZones} introduced spatial variation into this background field for the particular geometry of a grain-boundary depletion zone, using an assumed one-dimensional concentration profile characterized by a time-dependent depletion length. Such constructions are effective for the geometries for which they are derived, but do not provide a general representation of an evolving multidimensional solute field generated by arbitrary particle distributions and thermal histories.

Moreover, these long-wavelength variations need not equilibrate on the same timescale as the local diffusion fields surrounding individual precipitates. As the characteristic diffusion time increases with the square of the length scale, particle-scale concentration fields may be effectively quasi-steady while solute redistribution over larger, microstructurally relevant distances remains transient. A spatial precipitation model must therefore accommodate both the rapidly relaxing local diffusion fields that control individual particle growth and the slower evolution of the surrounding long-wavelength concentration field.

To address this limitation, a scale-separated monopole method is developed in which the local and long-range components of the concentration field are assigned different temporal descriptions. An Ewald decomposition of the periodic Green's function is used to define complementary spatial scales. Rapidly relaxing particle-scale diffusion fields are retained in analytical, quasi-steady monopole form, while the smooth long-wavelength concentration field is represented on a coarse spatial grid and evolved according to the transient diffusion equation. Growth and dissolution of the discrete precipitates provide conservative sources and sinks to this long-range field, which in turn determines the evolving local environment experienced by the particles. The resulting formulation retains the spatial distribution and diffusional interactions of individual precipitates while allowing solute redistribution over much larger distances to evolve on its physical diffusion timescale. Coupled with precipitation thermodynamics and classical nucleation theory, the method describes nucleation, growth, dissolution, and coarsening under general thermal histories. The formulation developed in this work is implemented in a precipitation-modelling code named Mono'Preci, the name used for the present calculations in the figures. The implementation is available through the repository identified in the Code availability statement.

The particle-scale monopole description is validated against analytical and transient finite-difference solutions and, at the population level, against a conventional KWN model using consistent thermodynamic and kinetic descriptions. The sensitivity of the scale-separated formulation to the parameters controlling the short-range/long-range decomposition is then assessed. Finally, the method is applied to precipitation within a transient concentration field generated self-consistently during a thermal history, demonstrating the coupling between local precipitation kinetics and mesoscale solute redistribution.

\section{The direct monopole method for precipitation kinetics}
\label{sec:direct_monopole}

\subsection{From Zener growth to interacting precipitate populations}

The direct monopole formulation used here follows the multi-particle diffusion treatment developed by Voorhees and Glicksman \cite{Voorhees1984,Voorhees1984SolutionSimulations} and subsequently generalized through a systematic multipole expansion by Akaiwa and Voorhees \cite{Akaiwa1994Late-stageFunctions}. Because this formulation is less commonly used in precipitation-kinetics modelling, we first give a brief synopsis of the underlying derivation. This also establishes the notation required for the scale-separated formulation introduced later.

The monopole description can be viewed as the lowest-order approximation within the broader class of boundary-source representations of diffusion around finite particles\cite{Jaswon1977IntegralEquationMethods}. In a general description, a multipole expansion represents a surface source distribution through successively higher-order angular contributions, the monopole being the leading spherically symmetric term. The monopole approximation therefore neglects angular concentration variations over the particle surface, and is appropriate when the particle radius is small relative to the characteristic length scale over which the external concentration field varies \cite{Akaiwa1994Late-stageFunctions}.

The connection between this approximation and conventional precipitation kinetics is most easily seen through the classical Zener description of diffusion-controlled growth \cite{Zener1949TheorySolution}. We initially consider a single solute and a spherical precipitate of radius $R_i$ centred at $\mathbf{R}_i$. When diffusion relaxes rapidly relative to motion of the precipitate interface, the surrounding concentration field is quasi-steady and satisfies

\begin{equation}
\nabla^2 c = 0
\label{eq:laplace}
\end{equation}

For an isolated particle in an infinite matrix,

\begin{equation}
c(\mathbf{x}) = c_\infty + (c_i-c_\infty) \frac{R_i}{|\mathbf{x}-\mathbf{R}_i|}
\label{eq:zener_profile}
\end{equation}

where $c_i$ and $c_\infty$ are the interfacial and far-field matrix compositions. The corresponding flux gives the Zener growth law

\begin{equation}
\frac{dR_i}{dt}
=
\frac{D}{R_i} \frac{c_\infty-c_i}{c_p-c_i}
\label{eq:zener_growth}
\end{equation}

where $D$ is the solute diffusivity in the matrix and $c_p$ is the precipitate composition.

Equation~\eqref{eq:zener_profile} may equivalently be written using the free-space Green's function,

\begin{equation}
c(\mathbf{x})
=
c_\infty + Q_i G(\mathbf{x}-\mathbf{R}_i) \qquad G(\mathbf{r}) = \frac{1}{4\pi|\mathbf{r}|}
\label{eq:single_monopole}
\end{equation}

where imposing $c=c_i$ at $r=R_i$ gives

\begin{equation}
Q_i = 4\pi R_i(c_i-c_\infty)
\label{eq:single_Q}
\end{equation}

Thus, for an isolated spherical precipitate in a uniform far field, the monopole representation reproduces the Zener concentration field exactly. With the present sign convention, $Q_i<0$ corresponds to growth and $Q_i>0$ to dissolution. Since $Q_i$ determines the integrated solute flux, the particle velocity is

\begin{equation}
\frac{dR_i}{dt} = -\frac{D Q_i} {4\pi R_i^2(c_p-c_i)}
\label{eq:Rdot_Q}
\end{equation}

which, together with Eq.~\eqref{eq:single_Q}, recovers Eq.~\eqref{eq:zener_growth}. 

For a population of particles, linearity permits the individual monopole fields to be superposed \cite{Voorhees1984SolutionSimulations}. In a periodic cell of volume $V$ containing $N$ precipitates,

\begin{equation}
c(\mathbf{x}) = \bar{c} + \sum_{j=1}^{N} Q_j G_{\mathrm{per}}(\mathbf{x}-\mathbf{R}_j)
\label{eq:periodic_superposition}
\end{equation}

where $\bar c$ is the spatially averaged matrix concentration determined from solute conservation and the zero-mean periodic Green's function satisfies

\begin{equation}
\nabla^2 G_{\mathrm{per}}(\mathbf{r}) = -\delta(\mathbf{r})+\frac{1}{V} \qquad \langle G_{\mathrm{per}}\rangle=0
\label{eq:periodic_green}
\end{equation}

The monopole strengths are coupled because the interfacial concentration of each particle contains contributions from all other particles. The boundary condition is applied to the surface-averaged concentration, retaining the spherically symmetric monopole contribution\cite{Voorhees1984SolutionSimulations}. The singular self-contribution is separated using

\begin{equation}
G_{\mathrm{per}}(\mathbf{r}) = \frac{1}{4\pi r} + G_0 + \mathcal{O}(r^2) \qquad r\rightarrow0
\label{eq:Gper_local}
\end{equation}

where

\begin{equation}
G_0 = \lim_{r\rightarrow0} \left[ G_{\mathrm{per}}(\mathbf{r}) - \frac{1}{4\pi r} \right]
\label{eq:G0}
\end{equation}

The interfacial condition for particle $i$ is therefore

\begin{equation}
c_i-\bar{c} = Q_i \left( \frac{1}{4\pi R_i}+G_0 \right) + \sum_{j\neq i} Q_j G_{\mathrm{per}}(\mathbf{R}_i-\mathbf{R}_j)
\label{eq:Qi_system}
\end{equation}

Defining the local background composition

\begin{equation}
c_i^{\mathrm{bg}} = \bar{c} + Q_iG_0 + \sum_{j\neq i} Q_j G_{\mathrm{per}}(\mathbf{R}_i-\mathbf{R}_j)
\label{eq:c_external}
\end{equation}

gives

\begin{equation}
Q_i = 4\pi R_i \left( c_i-c_i^{\mathrm{bg}} \right)
\label{eq:Qi_local_zener}
\end{equation}

Thus, each precipitate locally retains the form of the isolated Zener problem, but with the uniform far-field composition replaced by a local background determined self-consistently by the complete particle population.

Applying Eq.~\eqref{eq:Qi_system} to all particles gives

\begin{equation}
\mathbf{A}\mathbf{Q} = \mathbf{c}^{\,I} - \bar{c}\mathbf{1}
\label{eq:Q_matrix}
\end{equation}

where $\mathbf{Q}=(Q_1,\ldots,Q_N)^\mathrm{T}$, $\mathbf{c}^{\,I}=(c_1,\ldots,c_N)^\mathrm{T}$, and

\begin{equation}
A_{ij} =
\begin{cases}
\dfrac{1}{4\pi R_i}+G_0 & i=j \\[8pt]
G_{\mathrm{per}}(\mathbf{R}_i-\mathbf{R}_j) & i\neq j
\end{cases}
\label{eq:A_matrix}
\end{equation}

The interaction matrix is dense because every particle contributes to the local environment of every other particle. Solving Eq.~\eqref{eq:Q_matrix} determines the monopole strengths and, through Eq.~\eqref{eq:Rdot_Q}, the corresponding growth or dissolution rates. Diffusional interactions and soft impingement therefore emerge directly from the spatial particle configuration rather than through an additional mean-field correction.

\subsection{Coupling monopole transport to precipitation thermodynamics and kinetics}

The monopole formulation provides the solute fluxes to or from each explicitly represented precipitate. For multicomponent precipitation, a separate concentration field and monopole strength are associated with each independently diffusing component $\alpha$,

\begin{equation}
c^\alpha(\mathbf{x}) = \bar{c}^{\,\alpha} + \sum_{j=1}^{N} Q_j^\alpha G_{\mathrm{per}}(\mathbf{x}-\mathbf{R}_j)
\label{eq:multicomponent_concentration}
\end{equation}

while the particle geometry and Green-function interaction matrix are common to all components. Each particle is additionally assigned a precipitate phase, allowing multiple precipitate phases to coexist within the same calculation.

The monopole fluxes are coupled to conventional precipitation thermodynamics and kinetics \cite{Robson2004ModellingPrecipitation,Perez2008ImplementationPrecipitation,Du2012AAlloys}. For each particle, the interfacial compositions are determined by requiring local thermodynamic equilibrium, including the Gibbs--Thomson contribution, together with a common interface velocity for all components,

\begin{equation}
v_i^1 = v_i^2 = \cdots = v_i^S = \dot{R}_i
\label{eq:equal_velocity}
\end{equation}

The resulting coupled solution determines the component-specific monopole strengths $Q_i^\alpha$ and the growth or dissolution rate $\dot{R}_i$ of each particle, while global conservation is enforced independently for each solute species. Details of the thermodynamic relations, multicomponent growth solution, and solute-conservation procedure are given in Section S1 of the Supplementary Material.

New precipitates are introduced using classical nucleation theory (CNT), following the particle-based treatment of nucleation introduced by Sagui and co-workers \cite{Sagui1997NucleationState,Sagui1999TheorySeparation}. The important distinction from a mean-field precipitation model is that the nucleation driving force is evaluated from the local matrix composition. Consequently, the supersaturation and nucleation rate may vary spatially as the concentration fields generated by the existing precipitate population evolve. Candidate nuclei are introduced stochastically and subsequently evolve as members of the same explicit particle population. Details of the CNT formulation and stochastic nucleation procedure are given in Section S2 of the Supplementary Material.

\section{Validation and limitations of the direct monopole formulation}
\label{sec:direct_validation}

Before introducing the scale-separated formulation, we first establish the behaviour of the direct monopole method and the regime in which its quasi-steady treatment of diffusion is appropriate. All validation calculations are performed for precipitation of coherent Al$_3$Sc from an Al--Sc solid solution. This system provides a convenient model problem because precipitation can be described using a single independently diffusing solute while retaining the essential coupling between diffusion, interfacial thermodynamics, nucleation, growth, dissolution, and coarsening.

The material parameters used throughout the Al-Sc calculations are summarized in Table~\ref{tab:AlSc_parameters}. The parameterization is based on that employed by Eymann \textit{et al.}~\cite{Eymann2024AModel}, with the values used in the present calculations reported explicitly here. The same parameter set is used in the monopole, finite-difference, and KWN calculations unless otherwise stated. The Sc diffusivity is described by

\begin{equation}
D_{\mathrm{Sc}} = D_0\exp\left(-\frac{Q}{RT}\right)
\label{eq:AlSc_diffusivity}
\end{equation}

and the temperature-dependent Al$_3$Sc solubility product by

\begin{equation}
\log_{10}K(T) = -\frac{A}{T} + B + \frac{C}{T^2}
\label{eq:AlSc_solubility}
\end{equation}

\begin{table}[ht!]
\centering
\caption{Material parameters used for the Al-Sc calculations. The parameterization is based on that employed by Eymann \textit{et al.}~\cite{Eymann2024AModel}.}
\label{tab:AlSc_parameters}
\begin{tabular}{lll}
\hline
Parameter & Symbol & Value \\
\hline
Matrix atomic volume
& $v_m$
& $0.0166$~nm$^3$\,atom$^{-1}$ \\

Al$_3$Sc atomic volume
& $v_p$
& $0.0173$~nm$^3$\,atom$^{-1}$ \\

Matrix lattice parameter
& $a$
& $0.405$~nm \\

Interfacial energy
& $\gamma$
& $0.12$~J\,m$^{-2}$ \\

Diffusion prefactor
& $D_0$
& $5.3\times10^{-4}$~m$^2$\,s$^{-1}$ \\

Diffusion activation energy
& $Q$
& $174$~kJ\,mol$^{-1}$ \\

Solubility coefficient
& $A$
& $2576$~K \\

Solubility coefficient
& $B$
& $0.1158$ \\

Solubility coefficient
& $C$
& $-43018$~K$^2$ \\
\hline
\end{tabular}
\end{table}

First, a comparison is made of single-particle growth with transient finite-difference calculations, providing a direct test of the quasi-steady growth description. The finite-difference reference method is described in Section S3 of the Supplementary Material. We then compare complete precipitation kinetics with a conventional KWN model to determine whether the explicit interacting-particle formulation recovers the expected population-level behaviour. Additional finite-difference validation for interacting particle pairs is provided in Section S4 of the Supplementary Material.

\subsection{Single-particle growth}
\label{sec:single_particle_validation}

The first comparison is to the growth of a single Al$_3$Sc precipitate predicted by the direct formulation with a transient finite-difference calculation using identical thermodynamic and kinetic parameters. For an isolated spherical particle, the monopole formulation reduces analytically to the quasi-steady Zener solution, as shown in Sec.~\ref{sec:direct_monopole}. The comparison therefore primarily tests the direct formulation under finite-volume periodic conditions and the validity of the quasi-steady approximation relative to a transient diffusion calculation.

\begin{figure*}[ht!]
    \centering
    \includegraphics[width=0.92\textwidth]{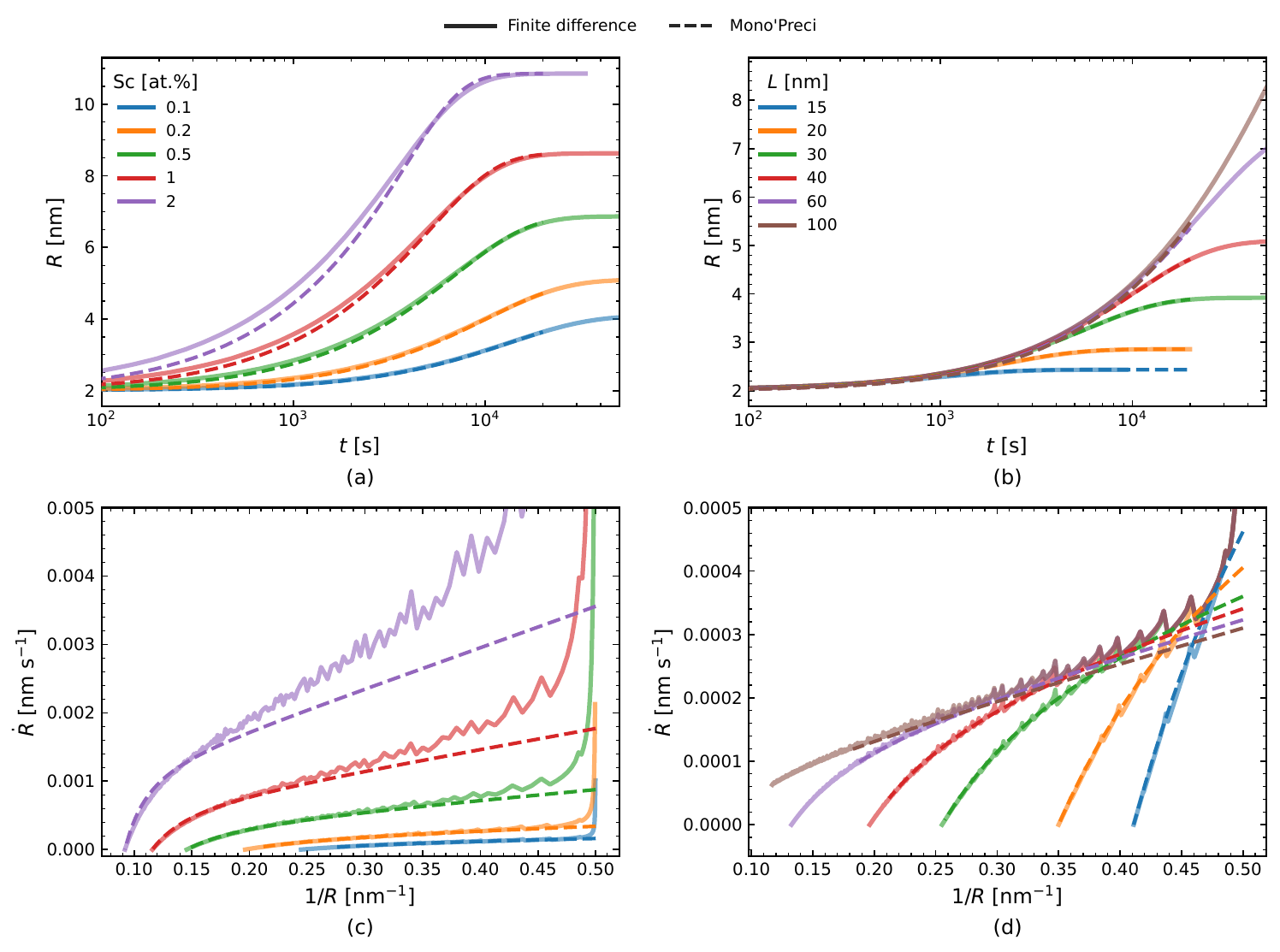}
    \caption{Validation of the direct monopole formulation for growth of a single Al$_3$Sc precipitate. (a) Particle radius as a function of time for different initial matrix Sc concentrations ($\ell = 40$~nm). (b) Particle radius as a function of time for different periodic cell dimensions $L$ (0.2at\%Sc). (c) Instantaneous growth rate $\dot{R}$ as a function of inverse particle radius for the composition series in (a).  (d) Corresponding growth-rate representation for the cell-size series in (b). Solid curves show the finite-difference calculations and dashed curves show the monopole calculations.}
    \label{fig:single_particle_validation}
\end{figure*}

Figure~\ref{fig:single_particle_validation}(a) compares the particle-radius histories for a range of initial Sc concentrations. The monopole calculations closely reproduce the finite-difference results over the complete composition range, including the increase in growth rate with supersaturation and the approach towards the finite-volume limiting particle radius. The corresponding cell-size series in Fig.~\ref{fig:single_particle_validation}(b) demonstrates that the periodic formulation also captures depletion of the finite solute reservoir as the simulation volume is reduced. The instantaneous growth rates in Figs.~\ref{fig:single_particle_validation}(c) and (d) provide a more sensitive comparison and reveal the largest differences during the initial rapid-growth regime.

These differences can be interpreted using the particle-growth P\'eclet number,

\begin{equation}
\mathrm{Pe}_R = \frac{R|\dot{R}|}{D}
\label{eq:particle_peclet}
\end{equation}

which is the ratio of the characteristic diffusion time, $R^2/D$, to the interface-motion time, $R/|\dot{R}|$. The quasi-steady limit therefore corresponds to $\mathrm{Pe}_R\ll1$.

\begin{figure*}[ht!]
    \centering
    \includegraphics[width=0.82\textwidth]{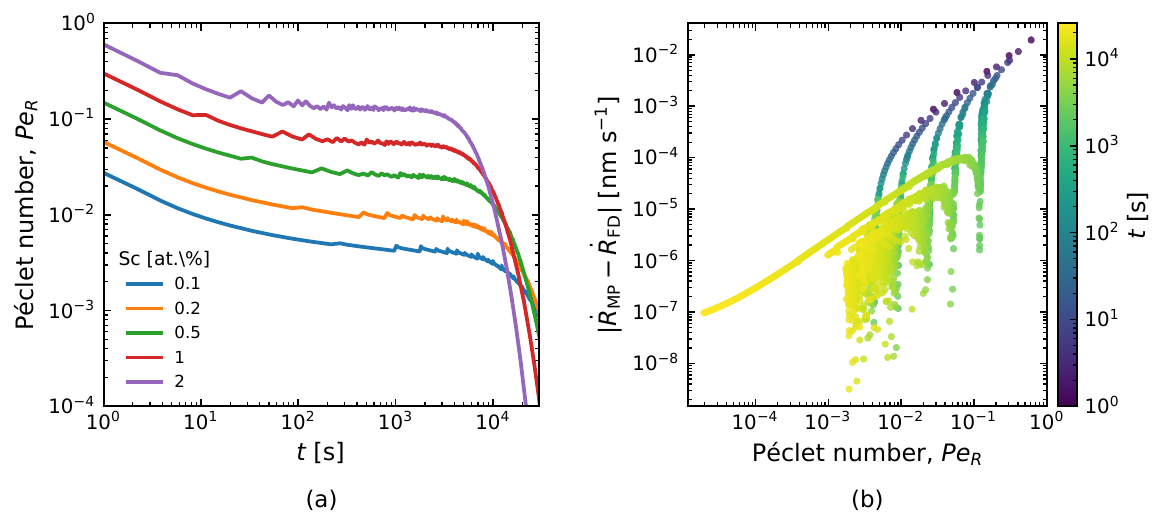}
    \caption{Relationship between the particle-growth P\'eclet number and the difference between the transient finite-difference and direct monopole descriptions for the single-particle composition series. (a) Evolution of the particle-growth P\'eclet number, $\mathrm{Pe}_R$, for initial matrix Sc concentrations of 0.1, 0.2, 0.5, 1, and 2~at.\%. (b) Absolute difference between the instantaneous particle growth rates predicted by the monopole and finite-difference models as a function of $\mathrm{Pe}_R$. The individual trajectories in (b) correspond to the compositions shown in (a) and shift towards higher $\mathrm{Pe}_R$ with increasing initial Sc concentration. Symbol colour indicates simulation time.}
    \label{fig:single_particle_peclet}
\end{figure*}

Figure~\ref{fig:single_particle_peclet}(a) shows that increasing supersaturation produces a pronounced increase in $\mathrm{Pe}_R$ during the initial rapid-growth regime. The relationship between this timescale separation and the remaining growth-rate discrepancy is shown in Fig.~\ref{fig:single_particle_peclet}(b). The largest differences between the transient finite-difference and direct monopole growth rates occur at early times, when $\mathrm{Pe}_R$ is largest, and decrease rapidly as the system approaches the quasi-steady regime.

The remaining discrepancy primarily reflects the comparison of a quasi-steady representation with a transient initial-value problem rather than an error introduced by the monopole approximation itself. For a single spherical particle, no higher-order angular contributions are required in the free-space quasi-steady limit. The finite-difference calculation develops the depletion field progressively from the prescribed initial matrix composition, whereas the direct monopole calculation establishes the quasi-steady field corresponding to the instantaneous particle state. The agreement therefore improves as diffusion becomes increasingly rapid relative to interface motion.

\subsection{Precipitate population kinetics and comparison with KWN}
\label{sec:kwn_validation}

Having established the behaviour of the direct formulation for an individual precipitate, we next examine whether the same method, when coupled to nucleation and precipitation thermodynamics, reproduces the evolution of a complete precipitate population. Monopole calculations were compared with a conventional Kampmann--Wagner numerical (KWN) model for precipitation of Al$_3$Sc from supersaturated Al--Sc solid solutions. The KWN calculations were performed using a Lagrangian implementation based on the PreciSo precipitation model developed by Perez and co-workers \cite{Perez2008ImplementationPrecipitation}. Identical thermodynamic and kinetic descriptions were used in the two calculations, including the matrix diffusivity, precipitate composition, interfacial energy, Gibbs--Thomson correction, and classical nucleation model. The principal distinction is therefore the representation of the precipitate population. KWN evolves a particle-size distribution within a spatially homogeneous matrix, whereas the monopole model represents individual precipitates whose growth and dissolution are coupled through their concentration fields.

\begin{figure*}[ht!]
    \centering
    \includegraphics[width=0.92\textwidth]{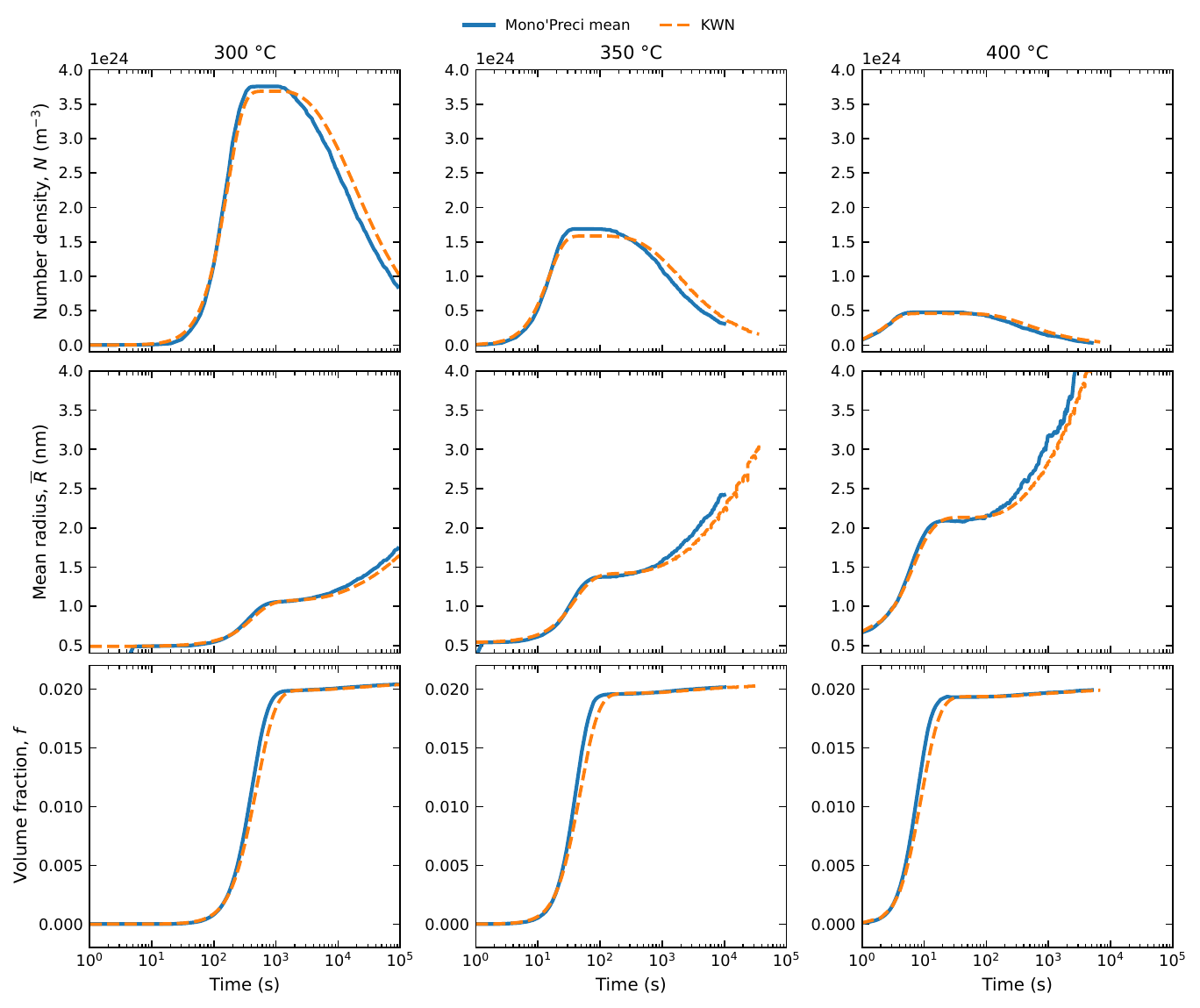}
    \caption{Comparison of Al$_3$Sc precipitation kinetics predicted by the monopole method and a KWN model at 300, 350, and 400~$^\circ$C. The upper, middle, and lower rows show precipitate number density $N$, mean radius $\overline{R}$, and precipitate volume fraction $f$, respectively. Dashed curves show the corresponding KWN calculations.}
    \label{fig:kwn_temperature_comparison}
\end{figure*}

Figure~\ref{fig:kwn_temperature_comparison} compares the precipitate number density $N$, mean radius $\overline{R}$, and volume fraction $f$ at 300, 350, and 400~$^\circ$C for an alloy containing Al--0.50~at.\% Sc.

The monopole calculations closely reproduce the KWN population kinetics across the three temperatures. Both descriptions capture the initial increase and subsequent decrease in number density, the continued increase in mean particle radius during coarsening, and the approach of the precipitate volume fraction towards its long-time value. The correspondence demonstrates that the discrete interacting-particle formulation recovers the expected population-level precipitation kinetics without imposing a spatially homogeneous matrix environment.

Small differences nevertheless remain, particularly during the nucleation stage. One source of these differences follows directly from the spatially resolved nature of the monopole calculations. In KWN, the nucleation rate is evaluated using a single homogeneous matrix composition. In the monopole model, the same CNT model can instead be evaluated using the local matrix composition at each candidate nucleation site, allowing concentration variations generated by the existing precipitates to feed back into subsequent nucleation.

\begin{figure*}[ht!]
    \centering
    \includegraphics[width=0.88\textwidth]{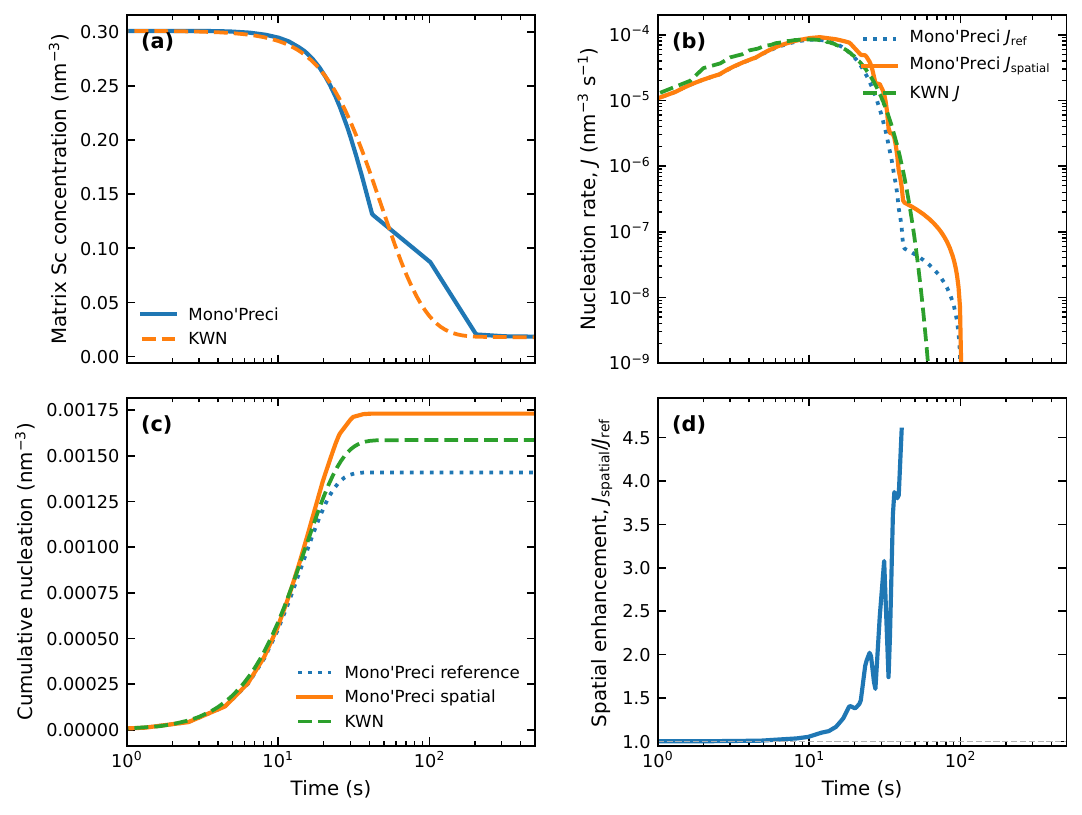}
    \caption{Origin of the difference in nucleation kinetics between monopole and KWN models at 350~$^\circ$C. (a) Evolution of the matrix Sc concentration. (b) Nucleation rate predicted by KWN and the corresponding monopole reference and spatially sampled rates. The reference rate $J_{\rm ref}$ is evaluated using the spatially averaged matrix composition, whereas $J_{\rm spatial}$ accounts for the local compositions sampled by candidate nucleation sites. (c) Corresponding cumulative number density of nucleated particles. (d) Ratio of the spatially sampled nucleation rate to the homogeneous reference value, $J_{\rm spatial}/J_{\rm ref}$.}
    \label{fig:kwn_nucleation}
\end{figure*}

This effect is examined at 350~$^\circ$C in Fig.~\ref{fig:kwn_nucleation}. To distinguish differences arising from the evolving mean matrix composition from those arising from spatial heterogeneity, two rates are evaluated from the monopole calculations. The reference rate, $J_{\rm ref}$, is evaluated using the spatially averaged matrix composition, whereas $J_{\rm spatial}$ accounts for the local compositions sampled during nucleation.

At early times, when the matrix remains nearly homogeneous, $J_{\rm spatial}$ and $J_{\rm ref}$ are similar. As the precipitate population develops, local depletion fields introduce spatial variations in composition and the two rates separate. This difference arises because the CNT nucleation rate depends nonlinearly on supersaturation, such that in general

\begin{equation}
\left\langle J\!\left(c(\mathbf{x})\right)\right\rangle
\neq
J\!\left(\left\langle c(\mathbf{x})\right\rangle\right)
\label{eq:spatial_nucleation_average}
\end{equation}

In the present calculations, spatial sampling gives $J_{\rm spatial}>J_{\rm ref}$ over part of the nucleation regime, resulting in a larger cumulative number of nucleation events. The reference rate evaluated from the spatially averaged composition remains close to the corresponding KWN rate. The difference therefore arises not from a different CNT model, but from applying the same composition-dependent nucleation kinetics to a spatially heterogeneous rather than homogeneous matrix.

The KWN comparison establishes that the direct monopole formulation recovers conventional population-level precipitation kinetics while retaining information about the local concentration environment of individual particles. Agreement in population averages, however, does not address a more fundamental limitation of the direct formulation. In the direct method the complete concentration field is assumed to respond quasi-statically to changes in the particle population. The consequences of this approximation depend on the spatial scale over which the concentration field varies and provide the motivation for the scale-separated formulation developed below.

\section{Scale-separated transient monopole formulation}
\label{sec:hybrid_formulation}

The direct monopole formulation performs well for precipitation in an initially homogeneous system, reproducing conventional population-level kinetics while retaining spatial correlations between explicitly represented precipitates. Its most significant approximation is that the complete concentration field is assumed to be quasi-steady.

The single-particle calculations in Sec.~\ref{sec:single_particle_validation} show that this approximation is well justified on the scale of an individual precipitate when the particle-growth P\'eclet number is small. Since the characteristic diffusion time over a length scale $\lambda$ scales as $\lambda^2/D$, concentration gradients on the particle scale may relax rapidly while longer-range variations extending over tens or hundreds of particle radii evolve on timescales comparable to the precipitation kinetics.

This distinction becomes important for spatially heterogeneous precipitation, where preferential nucleation, solute segregation, or other localized processes can generate composition variations over distances much larger than the precipitate dimensions. In the direct formulation, concentration variations at all wavelengths respond instantaneously to changes in the particle population. A fully transient treatment, such as finite difference or phase field, would remove this approximation, but would generally require particle-scale concentration variations to be resolved throughout the simulation domain.

\begin{figure*}[ht!]
    \centering
    \includegraphics[width=0.90\textwidth]{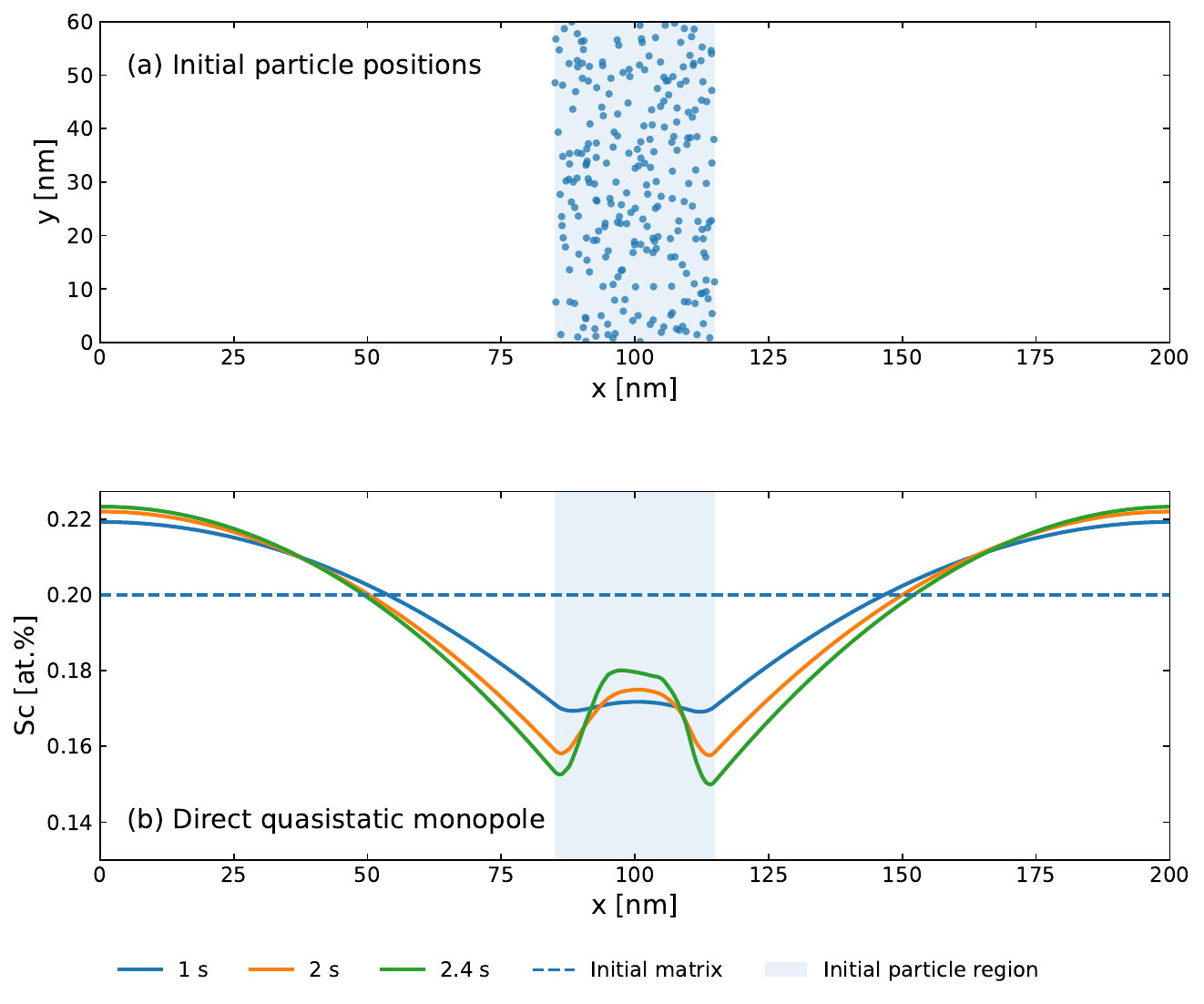}
    \caption{Illustration of the limitation of the direct quasistatic monopole formulation for spatially localized precipitates. (a) Initial positions of 250 Al$_3$Sc precipitates confined to a narrow region of an otherwise homogeneous Al--0.2~at.\% Sc matrix. (b) Matrix Sc concentration profiles predicted by the direct monopole formulation during growth of the localized particle population at $450\,^\circ$C. The shaded region indicates the initial particle-containing region and the dashed line the initial matrix composition.}
    \label{fig:direct_hybrid_motivation}
\end{figure*}

To illustrate the consequence of the globally quasi-steady approximation, we consider a deliberately simple spatially inhomogeneous precipitate population. Within a periodic box of dimensions $200\times60\times60$~nm, 250 identical Al$_3$Sc precipitates of initial radius $R=1$~nm were confined to a narrow region at the centre of an otherwise homogeneous Al--0.2~at.\% Sc matrix (Fig.~\ref{fig:direct_hybrid_motivation}(a)). The system was then held at $450\,^\circ$C.

The predicted concentration profiles, averaged over the $y$--$z$ plane, are shown in Fig.~\ref{fig:direct_hybrid_motivation}(b). Solute consumption produces the expected depletion within the particle-containing region, but the Sc concentration far from the precipitates immediately rises significantly above the initial matrix composition. Since the matrix is initially uniform and the growing precipitates act only as sinks for Sc, this nonphysical far-field enrichment results from imposing the periodic quasi-steady solution corresponding to the instantaneous particle state.

In a periodic domain, a concentration Fourier mode of wavelength $\lambda$ relaxes with characteristic time $\tau_\lambda=\lambda^2/(4\pi^2D)$. For the longest non-zero mode in the present cell, $\lambda=L_x=200$~nm. At $450\,^\circ$C, $D_{\mathrm{Sc}}=143$~nm$^2$\,s$^{-1}$, giving $\tau_L\simeq7.1$~s, approximately three times the 2.4~s interval shown in Fig.~\ref{fig:direct_hybrid_motivation}(b). The direct formulation nevertheless places this long-wavelength component at its quasi-steady value at every instant. The nonphysical far-field enrichment is therefore a direct manifestation of the globally quasi-steady approximation.

The periodic Green's function provides a natural means of separating fast near-field from slow far-field solute relaxation. Its real-space lattice sum converges only slowly and is commonly evaluated using an Ewald representation \cite{Ewald1921,FrenkelSmit}. The Ewald construction decomposes the periodic Green's function into complementary short- and long-range contributions,

\begin{equation}
G_{\mathrm{per}}(\mathbf{r}) = G_{\mathrm{SR}}(\mathbf{r};\ell) + G_{\mathrm{LR}}(\mathbf{r};\ell)
\label{eq:ewald_split}
\end{equation}

where the matching length $\ell$ controls the spatial scale of the separation. For the Gaussian Ewald splitting used here \cite{Ewald1921,FrenkelSmit}, the short-range contribution is

\begin{equation}
G_{\mathrm{SR}}(\mathbf{r};\ell) = \frac{1}{4\pi r}\operatorname{erfc}\left(\frac{r}{\ell}\right) - \frac{\ell^2}{4V}
\label{eq:G_SR}
\end{equation}

where the $\operatorname{erfc}(r/\ell)$ factor screens the long-range $1/r$ interaction while leaving its singular near-field behaviour unchanged as $r\rightarrow0$. The spatially varying part of $G_{\mathrm{SR}}$ therefore decays rapidly for $r\gtrsim\ell$. The complementary contribution,

\begin{equation}
G_{\mathrm{LR}}(\mathbf{r};\ell) = G_{\mathrm{per}}(\mathbf{r}) - G_{\mathrm{SR}}(\mathbf{r};\ell)
\label{eq:G_LR}
\end{equation}

is smooth on the particle scale.

In the conventional Ewald construction this decomposition is purely mathematical; the two contributions together reconstruct the same instantaneous quasi-steady periodic Green's function. Here the decomposition is instead used to separate the diffusion dynamics. The short-range contribution remains quasi-steady and tied to the instantaneous particle monopole strengths, whereas the complementary long-range contribution is represented by an independently evolving transient field $C_{\mathrm{LR}}(\mathbf{x},t)$. The matrix concentration is therefore represented as

\begin{equation}
c(\mathbf{x},t) = C_{\mathrm{LR}}(\mathbf{x},t) + \sum_{j=1}^{N} Q_j(t)G_{\mathrm{SR}}\left(\mathbf{x}-\mathbf{R}_j;\ell\right)
\label{eq:hybrid_concentration}
\end{equation}

The central approximation is therefore not that diffusion is quasistatic, but that diffusion is quasistatic only over the short length scales represented by $G_{\mathrm{SR}}$. The matching length $\ell$ defines a smooth transition between the quasistatic and transient descriptions rather than a physical boundary between two diffusion domains.

\subsection{Short-range particle coupling}

For a known long-range field, $C_{\mathrm{LR}}$ acts as the local background concentration for the short-range monopole problem. The interfacial condition for particle $i$ is

\begin{equation}
c_i = C_{\mathrm{LR}}(\mathbf{R}_i,t) + \sum_{j=1}^{N} A_{ij}^{\mathrm{SR}}Q_j
\label{eq:hybrid_interface_condition}
\end{equation}

where $c_i$ is the interfacial matrix concentration determined from the conditions described in Sec.~\ref{sec:direct_monopole}. The short-range interaction operator is

\begin{equation}
\mathbf{A}^{\mathrm{SR}} = \mathbf{B} - \frac{\ell^2}{4V}\mathbf{1}\mathbf{1}^{T}
\label{eq:short_range_operator}
\end{equation}

where, for $i\neq j$,

\begin{equation}
B_{ij} = \frac{\operatorname{erfc}(\alpha r_{ij})}{4\pi r_{ij}},
\label{eq:short_range_pair}
\end{equation}

with $r_{ij}$ the minimum-image separation, and

\begin{equation}
B_{ii} = \frac{1}{4\pi R_i} - \frac{1}{2\pi^{3/2}\ell}
\label{eq:short_range_self}
\end{equation}

The diagonal term combines the finite-radius monopole self contribution with the finite correction introduced by Ewald screening, while the uniform correction in Eq.~\eqref{eq:short_range_operator} ensures consistency with the zero-mean periodic Green's function. The monopole strengths are obtained from

\begin{equation}
\mathbf{A}^{\mathrm{SR}}\mathbf{Q} = \mathbf{c}_{\Gamma} - \mathbf{C}_{\mathrm{LR}}
\label{eq:hybrid_Q_system}
\end{equation}

where $\mathbf{C}_{\mathrm{LR}}$ denotes the long-range field evaluated at the particle positions.

Because the screened spatially varying interaction decays rapidly with separation, it is evaluated only within a finite cutoff distance $r_{\mathrm{cut}}$. Here, $r_{\mathrm{cut}}=3\ell$ is used, beyond which the remaining screened interaction is negligible. Each particle therefore couples directly only to a local neighbourhood, while more distant interactions are mediated through the transient long-range field. The dense interaction matrix of the direct formulation is thereby replaced by a sparse short-range matrix. Additional numerical implementation details are given in Section S5 of the Supplementary Material.

\subsection{Transient evolution of the long-range concentration field}

Let $M_i^\alpha$ denote the number of atoms of component $\alpha$ contained in particle $i$. Growth or dissolution produces an equal and opposite change in the matrix solute inventory, so the long-range field obeys

\begin{equation}
\frac{\partial C_{\mathrm{LR}}^\alpha}{\partial t} = D^\alpha\nabla^2C_{\mathrm{LR}}^\alpha - \sum_i\dot{M}_i^\alpha\rho_\ell(\mathbf{x}-\mathbf{R}_i)
\label{eq:LR_diffusion_source}
\end{equation}

with an independent field evolved for each diffusing component $\alpha$.

The spatial form of the source follows from the same Gaussian Ewald decomposition used to define the short-range Green's function,

\begin{equation}
\delta(\mathbf{r}) = \left[\delta(\mathbf{r})-\rho_\ell(\mathbf{r})\right] + \rho_\ell(\mathbf{r})
\label{eq:ewald_source_split}
\end{equation}

where

\begin{equation}
\rho_\ell(\mathbf{r}) = \frac{1}{\pi^{3/2}\ell^3} \exp\left(-\frac{r^2}{\ell^2}\right)
\label{eq:ewald_gaussian}
\end{equation}

is a normalized Gaussian screening distribution. The first term generates the screened spatially varying part of the short-range Green's function, while the second provides the complementary smooth source for the long-range field. Its Fourier transform is

\begin{equation}
W_\ell(k)=\widehat{\rho}_\ell(\mathbf{k}) = \exp\left(-\frac{k^2\ell^2}{4}\right)
\label{eq:ewald_filter}
\end{equation}

Thus, the Gaussian distribution of solute transferred to the long-range field follows directly from the Ewald decomposition rather than representing an additional numerical smoothing operation.

Fourier transformation of Eq.~\eqref{eq:LR_diffusion_source} gives

\begin{equation}
\frac{\partial\widehat{C}_{\mathrm{LR}}^\alpha(\mathbf{k},t)}{\partial t} = -D^\alpha k^2\widehat{C}_{\mathrm{LR}}^\alpha(\mathbf{k},t) - W_\ell(k)\sum_i\dot{M}_i^\alpha(t)\exp(-i\mathbf{k}\cdot\mathbf{R}_i)
\label{eq:LR_fourier_evolution}
\end{equation}

In the numerical scheme, solute exchange with the particles is separated from transient long-range diffusion. Over a particle update of duration $\Delta t$, the short-range calculation determines the finite particle inventory change $\Delta M_i^\alpha$, which is transferred to the long-range field according to

\begin{equation}
\Delta C_{\mathrm{LR}}^\alpha(\mathbf{x}) = -\sum_i\Delta M_i^\alpha\rho_\ell(\mathbf{x}-\mathbf{R}_i)
\label{eq:LR_inventory_transfer}
\end{equation}

Because $\rho_\ell$ is normalized, this transfers exactly the solute gained or lost by the particle population and preserves the total solute inventory.

In the absence of a particle source, each Fourier mode evolves exactly over $\Delta t$ according to

\begin{equation}
\widehat{C}_{\mathrm{LR}}^\alpha(\mathbf{k},t+\Delta t) = \exp(-D^\alpha k^2\Delta t) \widehat{C}_{\mathrm{LR}}^\alpha(\mathbf{k},t)
\label{eq:LR_diffusion_propagator}
\end{equation}

Diffusion and solute transfer are combined symmetrically, i.e. the long-range field is first diffused over $\Delta t/2$, the particle inventory change is applied through Eq.~\eqref{eq:LR_inventory_transfer}, and the field is then diffused over the remaining $\Delta t/2$. The zero Fourier mode is unchanged by diffusion and evolves only through solute exchange with the particle population, thereby determining the spatial mean of the long-range field. Details of the conservative particle-to-grid transfer and numerical implementation are given in Section S5 of the Supplementary Material.

\subsection{Recovery of transient long-range diffusion}
\label{sec:hybrid_demonstration}

\begin{figure*}[ht!]
    \centering
    \includegraphics[width=0.90\textwidth]{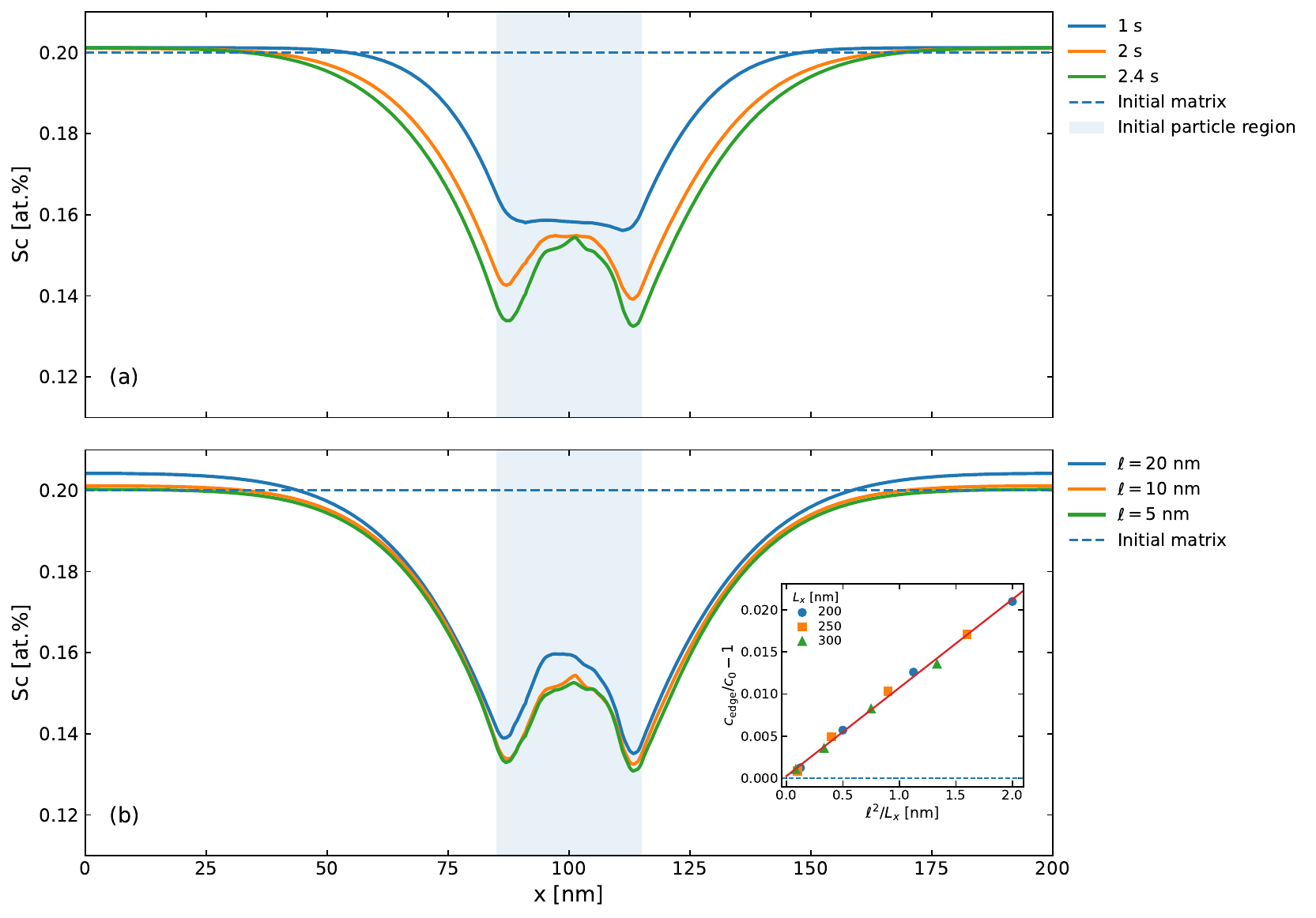}
    \caption{Transient Sc concentration profiles predicted by the scale-separated formulation for the spatially localized Al$_3$Sc particle population introduced in Fig.~\ref{fig:direct_hybrid_motivation}. (a) Evolution of the concentration field for $\ell=10$~nm at $450\,^\circ$C. (b) Concentration profiles at $t=2.4$~s for matching lengths $\ell=5$, 10, and 20~nm. The shaded region indicates the initial particle-containing region and the dashed line the initial matrix composition. The inset shows the normalized far-field excess concentration, $c_{\mathrm{edge}}/c_0-1$, as a function of $\ell^2/L_x$ for $L_x=200$, 250, and 300~nm, where $c_{\mathrm{edge}}$ is averaged within 10~nm of the periodic boundaries.}
    \label{fig:hybrid_transient_profiles}
\end{figure*}

We now return to the spatially localized particle population introduced in Fig.~\ref{fig:direct_hybrid_motivation}. Figure~\ref{fig:hybrid_transient_profiles} shows the corresponding behaviour obtained with the scale-separated formulation.

In contrast to the direct solution, the matrix far from the localized particle population remains close to its initial composition at early times. Solute depletion is initially concentrated around the growing precipitates and broadens progressively as the long-range field evolves by transient diffusion (Fig.~\ref{fig:hybrid_transient_profiles}(a)). A small residual far-field response remains at finite matching length.

Figure~\ref{fig:hybrid_transient_profiles}(b) shows that this residual response depends on $\ell$. For $\ell=20$~nm, the profile exhibits a shallower depletion region and greater remote elevation than for the smaller matching lengths. Reducing $\ell$ to 10~nm substantially reduces these differences, while the profiles for $\ell=10$ and 5~nm are nearly coincident.

The origin of the residual far-field elevation follows directly from the short-range contribution. Substituting Eq.~\eqref{eq:G_SR} gives

\begin{equation}
c_{\mathrm{SR}}(\mathbf{x},t) = \sum_iQ_i(t)\frac{\operatorname{erfc}(r_i/\ell)}{4\pi r_i} -
\frac{\ell^2}{4V}\sum_iQ_i(t)
\label{eq:c_SR_expanded}
\end{equation}

where $r_i=|\mathbf{x}-\mathbf{R}_i|$. Far from the localized particle population, the screened contributions decay and

\begin{equation}
c_{\mathrm{SR}}^{\mathrm{far}} \simeq-\frac{\ell^2}{4V}\sum_iQ_i
\label{eq:SR_far_uniform}
\end{equation}

The remaining contribution is uniform throughout the periodic domain and depends on the net monopole strength. Since $\sum_iQ_i<0$ during growth of the localized particle population, it is positive and produces the residual far-field elevation in Fig.~\ref{fig:hybrid_transient_profiles}(b).

For the calculations shown in the inset, the transverse dimensions and localized particle population are fixed while $L_x$ is varied. Since $V=L_xL_yL_z$, Eq.~\eqref{eq:SR_far_uniform} predicts approximately

\begin{equation}
c_{\mathrm{SR}}^{\mathrm{far}}\propto\frac{\ell^2}{L_x}
\label{eq:SR_far_scaling}
\end{equation}

provided that the net monopole strength does not change substantially. The calculations for three domain lengths and four matching lengths collapse approximately onto this expected dependence.

\begin{figure*}[ht!]
    \centering
    \includegraphics[width=0.88\textwidth]{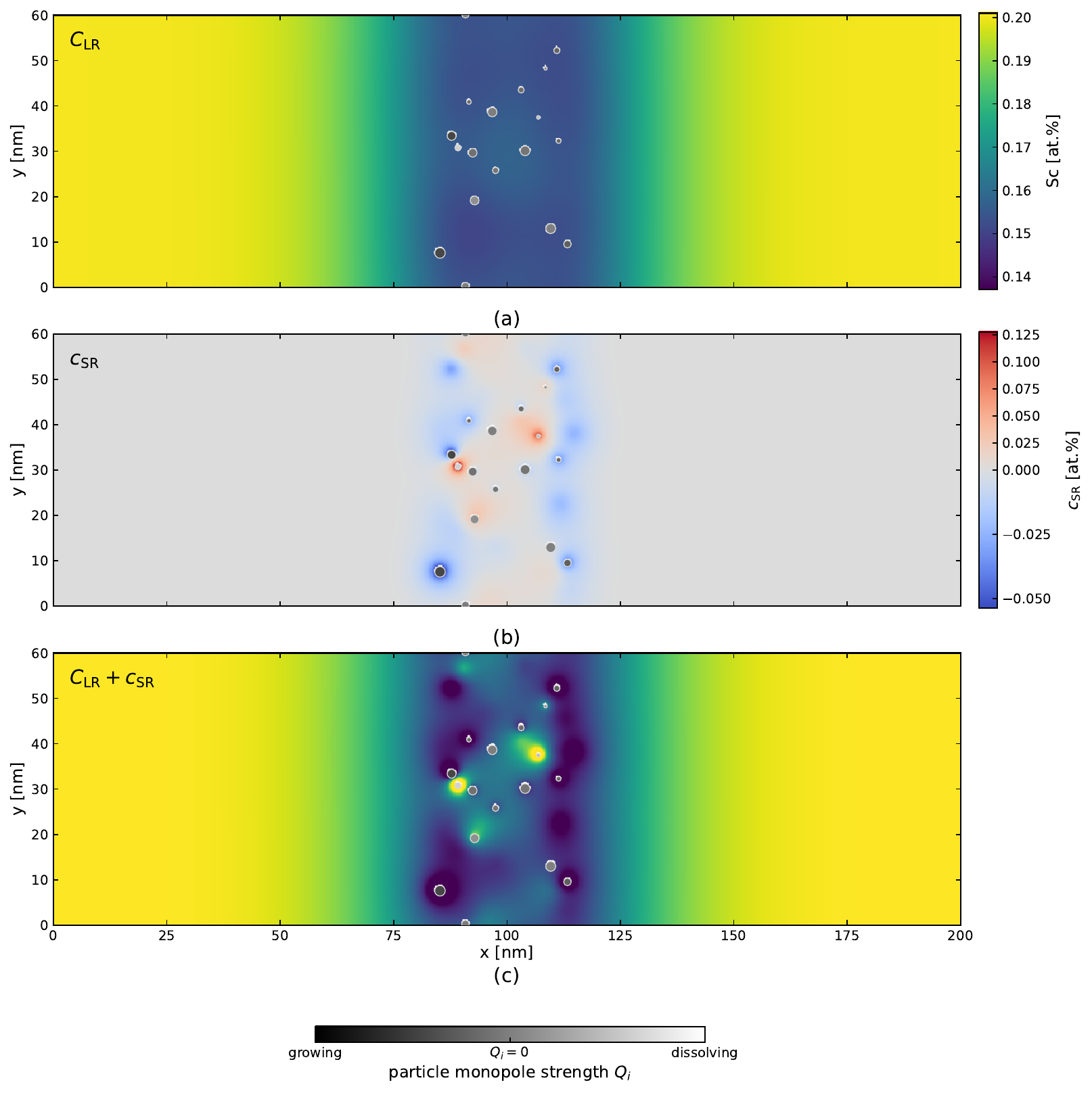}
    \caption{Spatial decomposition of the Sc concentration field in the scale-separated formulation for $\ell=10$~nm. (a) Smooth long-range field $C_{\mathrm{LR}}$. (b) Short-range contribution $c_{\mathrm{SR}}=\sum_jQ_jG_{\mathrm{SR}}$, containing the localized particle-scale concentration variations and the uniform periodic contribution. (c) Reconstructed matrix concentration $C_{\mathrm{LR}}+c_{\mathrm{SR}}$. Particle symbols indicate the instantaneous monopole strengths $Q_j$, with growing particles acting as solute sinks and dissolving particles as solute sources.}
    \label{fig:hybrid_field_decomposition}
\end{figure*}

The spatial character of the decomposition is illustrated in Fig.~\ref{fig:hybrid_field_decomposition} for $\ell=10$~nm. The long-range field is smooth on the particle scale and contains the collective depletion field, whereas the short-range contribution contains the localized particle-scale variations together with the uniform periodic contribution. Their superposition gives the complete matrix concentration.

The decomposition does not introduce a sharp boundary between short- and long-range diffusion. Instead, $\ell$ defines a smooth matching between the quasistatic particle-scale field and the transient long-range field. Its appropriate value is therefore determined by the physical timescale separation together with particle-scale and numerical-resolution constraints.

\subsection{Selection of the matching length}
\label{sec:matching_length}

Whether a transient treatment is required depends not only on the relaxation time of a spatial mode but also on the extent to which that mode is excited by the evolving particle population. A Fourier mode with wavevector $\mathbf{k}$ relaxes on the timescale $\tau_k^\alpha=(D^\alpha k^2)^{-1}$. Although low-$k$ modes relax slowly, their amplitudes are small for a statistically homogeneous precipitate population, helping to explain the success of the globally quasistatic formulation in Sec.~\ref{sec:kwn_validation}. Spatially heterogeneous particle populations can instead generate substantial low-$k$ components, which must retain their transient diffusion dynamics.

The matching length $\ell$ determines the scale over which the particle-generated response is retained in the quasistatic short-range field. A useful local criterion follows by comparing the diffusion time over this scale, $\tau_\ell^\alpha\sim\ell^2/D^\alpha$, with the timescale of particle evolution, $\tau_{R,i}\sim R_i/|\dot R_i|$. Requiring $\tau_\ell^\alpha\ll\tau_{R,i}$ gives

\begin{equation}
\frac{\tau_\ell^\alpha}{\tau_{R,i}} \sim \frac{\ell^2|\dot R_i|}{D^\alpha R_i} \ll1
\label{eq:matching_timescale_criterion}
\end{equation}

Introducing the particle-scale P\'eclet number $\mathrm{Pe}_{R,i}^{\alpha}=R_i|\dot R_i|/D^\alpha$ gives

\begin{equation}
\mathrm{Pe}_{R,i}^{\alpha} \left(\frac{\ell}{R_i}\right)^2 \ll1 
\label{eq:matching_Pe_criterion}
\end{equation}

This is a conservative local criterion for extending the quasistatic particle response to distances of order $\ell$. It generalizes the particle-scale condition $\mathrm{Pe}_{R,i}^{\alpha}\ll1$ established in Sec.~\ref{sec:single_particle_validation}: extending the quasistatic description from $R_i$ to $\ell$ increases the characteristic diffusion time by $(\ell/R_i)^2$.

The matching length cannot be reduced arbitrarily. Decreasing $\ell$ transfers progressively shorter-wavelength variations to the gridded long-range field, which must resolve the matching scale. The relevant numerical parameter is therefore $\Delta x/\ell$. For the calculations in Fig.~\ref{fig:hybrid_transient_profiles}, $\Delta x=2$~nm, giving $\Delta x/\ell=0.10$, 0.20, and 0.40 for $\ell=20$, 10, and 5~nm, respectively. The $\ell=5$~nm case was additionally repeated with $\Delta x=1$~nm, with essentially unchanged far-field concentrations and concentration profiles. The observed matching-length dependence is therefore not attributable to inadequate long-range grid resolution. Additional sensitivity tests are presented in Section S6 of the Supplementary Material.

The lower useful range of $\ell$ is also constrained by the particle radius. The hybrid construction requires a region outside each particle over which the screened quasistatic monopole field can be represented, so $R_i/\ell<1$ is desirable. For the screened Green's function used here, the finite-radius contribution to the self coefficient in Eq.~\eqref{eq:short_range_self} exceeds the Ewald self correction when

\begin{equation}
\frac{R_i}{\ell}<\frac{\sqrt{\pi}}{2}\simeq0.886
\label{eq:R_over_ell}
\end{equation}

This is not a fundamental physical boundary, but provides a useful diagnostic of the separation between the particle and matching scales.

The useful range of matching lengths is therefore approximately

\begin{equation}
R_i\lesssim\ell\ll\sqrt{D^\alpha\tau_{R,i}} \qquad \Delta x\ll\ell
\label{eq:matching_window}
\end{equation}

or equivalently,

\begin{equation}
1\lesssim\frac{\ell}{R_i} \ll \frac{1}{\sqrt{\mathrm{Pe}_{R,i}^{\alpha}}}
\label{eq:matching_window_Pe}
\end{equation}

A scale-separated regime exists when particle-scale diffusion is sufficiently rapid to provide an intermediate range between the particle radius and the length scale over which transient diffusion becomes important. Within this range, predictions should become weakly dependent on the precise choice of $\ell$, subject to adequate resolution of the long-range field.

\section{Precipitation in a spatially heterogeneous solute field}
\label{sec:heterogeneous_precipitation}

The principal motivation for the scale-separated formulation is to describe precipitation when the evolving microstructure generates composition variations over length scales substantially larger than the individual precipitates. To demonstrate this capability, we consider a model problem in which the spatial heterogeneity develops from an initially uniform matrix. A single pre-existing Al$_3$Sc precipitate of radius $R_0=2$~nm is placed at the centre of an Al--0.25~at.\% Sc matrix. The alloy is cooled from $550\,^\circ$C to $350\,^\circ$C at $5\,^\circ$C\,s$^{-1}$ and subsequently held at $350\,^\circ$C, with homogeneous nucleation enabled throughout. Cooling is sufficiently rapid to suppress secondary nucleation during this stage while allowing the initial precipitate to grow and generate a long-range Sc-depleted region. Secondary precipitates formed during the subsequent hold therefore nucleate and grow within a transient composition field generated self-consistently by the evolving microstructure.

The calculation was performed in a cubic periodic domain of side length $500$~nm using a matching length $\ell=40$~nm, a short-range cutoff $r_{\mathrm{cut}}=120$~nm, and a long-range grid spacing $\Delta x=5$~nm (Table~\ref{tab:heterogeneous_case}). The largest precipitate reached approximately $23.3$~nm, corresponding to $R_{\max}/\ell\simeq0.58$. Calculations with $\ell=30$--60~nm showed only weak sensitivity of the global precipitation kinetics and concentration profiles to the matching length, apart from fluctuations associated with stochastic nucleation. A separate calculation in an $800$~nm cubic domain confirmed that the long-range depletion field was sufficiently contained within the $500$~nm production domain. Matching-length and long-range grid-resolution sensitivity tests are presented in Section S6 of the Supplementary Material.

\begin{table}[t]
\centering
\caption{Parameters used for the spatially heterogeneous Al--Sc precipitation calculation.}
\label{tab:heterogeneous_case}
\begin{tabular}{lll}
\hline
Parameter & Symbol & Value \\
\hline
Simulation domain & $L_x\times L_y\times L_z$ & $500\times500\times500$~nm$^3$ \\
Initial matrix Sc concentration & $c_{\mathrm{Sc},0}$ & 0.25~at.\% \\
Initial temperature & $T_0$ & $550\,^\circ$C \\
Holding temperature & $T_{\mathrm{hold}}$ & $350\,^\circ$C \\
Cooling rate & $\dot T$ & $-5\,^\circ$C\,s$^{-1}$ \\
Initial seed radius & $R_0$ & 2.0~nm \\
Hybrid matching length & $\ell$ & 40~nm \\
Short-range cutoff & $r_{\mathrm{cut}}$ & 120~nm \\
Long-range grid spacing & $\Delta x$ & 5~nm \\
\hline
\end{tabular}
\end{table}

\begin{figure}[t]
    \centering
    \includegraphics[width=\linewidth]{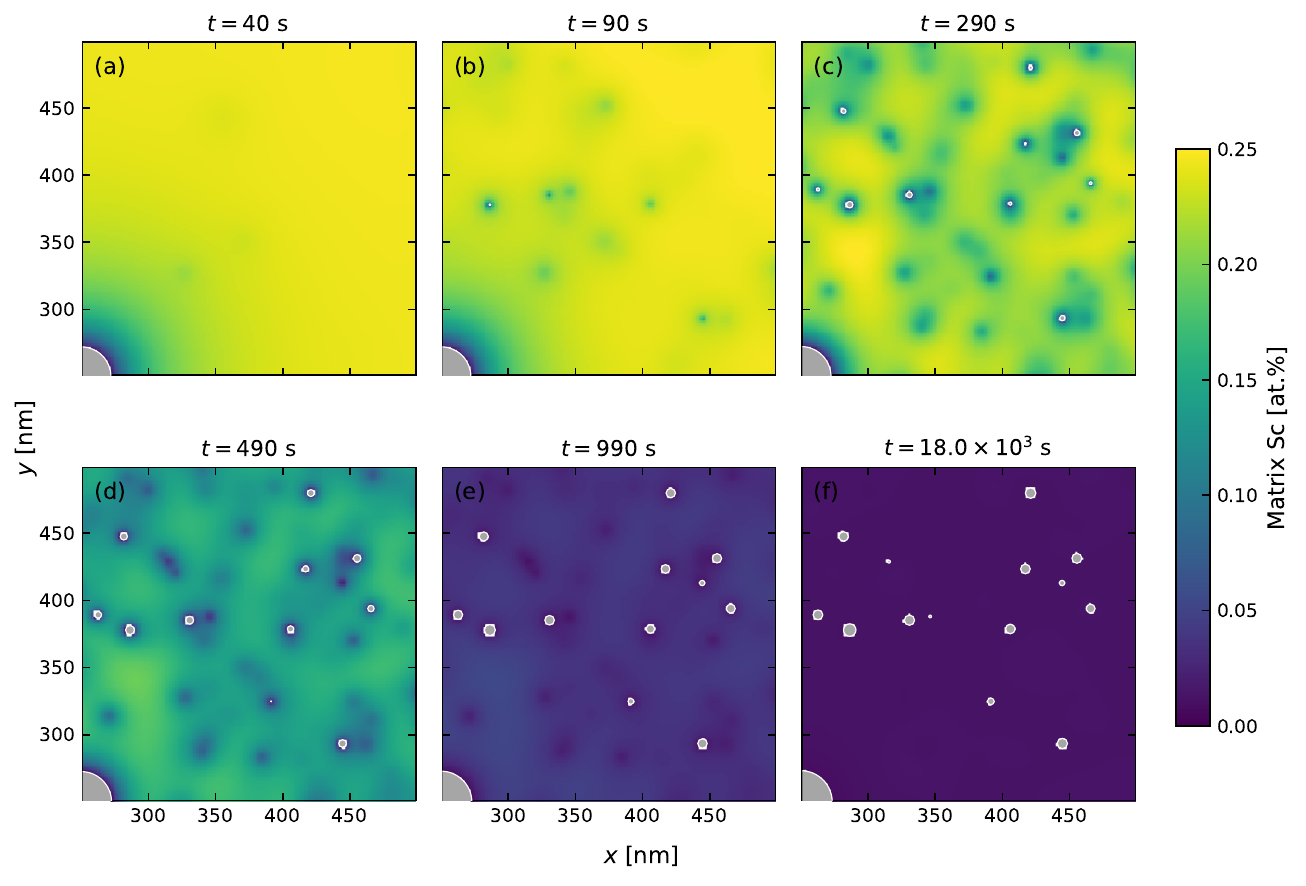}
    \caption{Evolution of the matrix Sc concentration during precipitation in an initially homogeneous Al--0.25~at.\% Sc alloy containing a single pre-existing central Al$_3$Sc precipitate. The alloy was cooled from $550\,^\circ$C to $350\,^\circ$C at $5\,^\circ$C\,s$^{-1}$ and subsequently held at $350\,^\circ$C. The panels show a quarter of the plane passing through the original particle at (a) 40~s, (b) 90~s, (c) 290~s, (d) 490~s, (e) 990~s, and (f) $18.0\times10^3$~s. Grey regions indicate cross-sections of precipitates intersecting the displayed plane. A common concentration scale is used for all panels.}
    \label{fig:heterogeneous_fields}
\end{figure}

Figure~\ref{fig:heterogeneous_fields} shows the evolution of the matrix Sc concentration in a plane through the original precipitate. At $t=40$~s, approximately the end of cooling, growth of the seed has generated a pronounced depletion field extending well beyond the particle, while the more distant matrix remains close to its initial composition. Secondary precipitation subsequently develops preferentially outside this depleted region. By $t=290$--490~s, a large population of secondary Al$_3$Sc precipitates has formed, producing increasingly complex local concentration fields. Continued precipitation reduces the matrix Sc concentration throughout the domain and, by $t=18.0\times10^3$~s, the long-wavelength concentration variation has largely relaxed while the spatially heterogeneous precipitate population remains.

\begin{figure}[t]
    \centering
    \includegraphics[width=\linewidth]{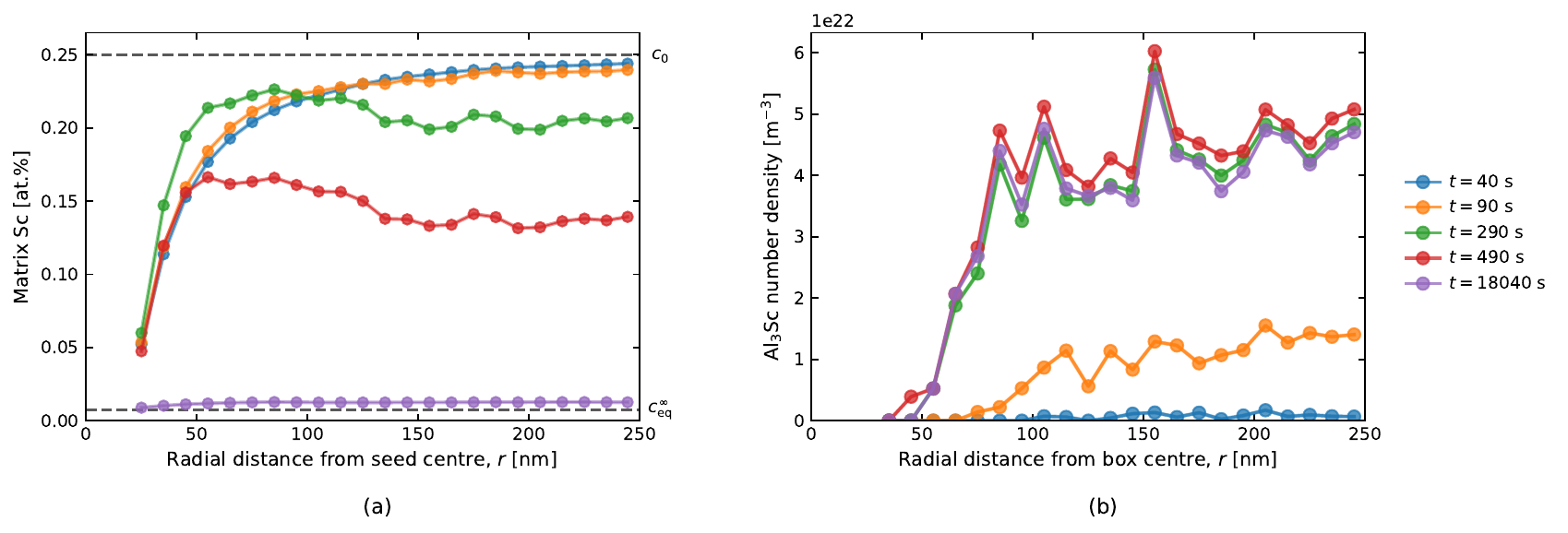}
    \caption{Radial evolution of (a) matrix Sc concentration and (b) secondary Al$_3$Sc number density about the centre of the original seed particle. The original seed is excluded from the particle-number statistics. The dashed line in (a) indicates the initial matrix composition, $c_0=0.25$~at.\% Sc, and the lower reference line indicates the planar equilibrium matrix composition at $350\,^\circ$C. Number densities in (b) are calculated from concentric spherical shells lying outside the instantaneous seed particle.}
    \label{fig:heterogeneous_radial}
\end{figure}

The coupling between the transient composition field and secondary precipitation is quantified in Fig.~\ref{fig:heterogeneous_radial}. The radial concentration profiles show the depletion field generated by the original seed and its subsequent evolution as secondary precipitation develops. Because the homogeneous nucleation rate depends strongly on local supersaturation, secondary nucleation is suppressed near the seed and increases with radial distance. By $t=290$--490~s, the outer matrix reaches a secondary precipitate number density of approximately $5\times10^{22}$~m$^{-3}$, while the population within approximately 100~nm of the seed remains substantially lower.

\begin{figure}[t]
    \centering
    \includegraphics[width=\linewidth]{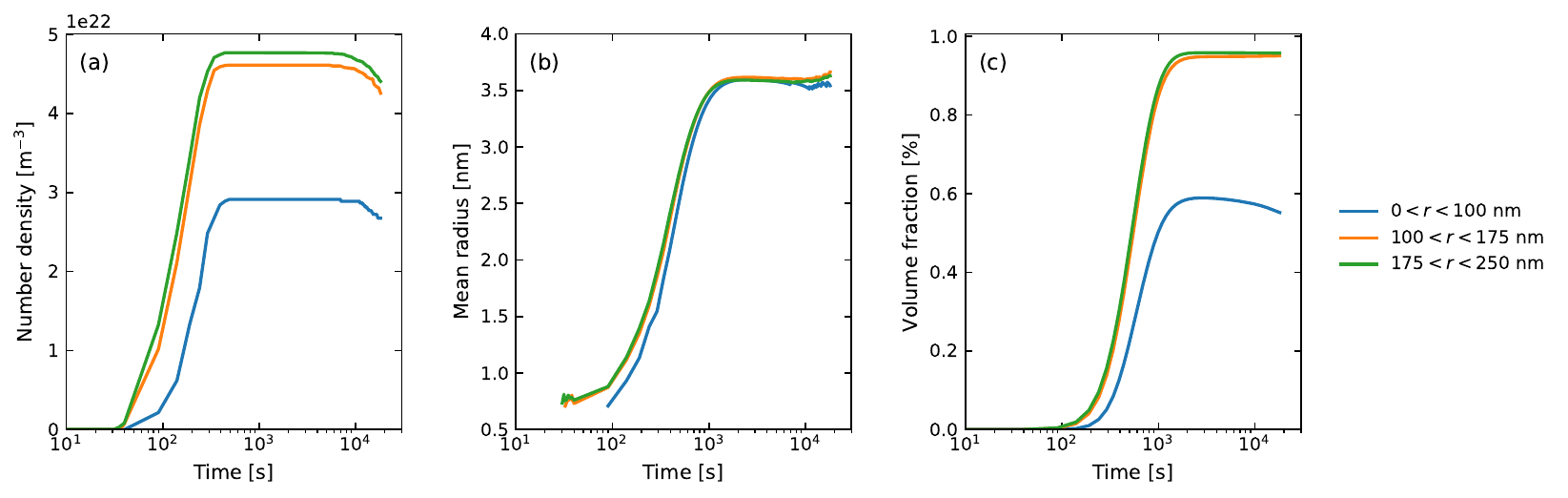}
    \caption{Spatially resolved precipitation kinetics for the secondary Al$_3$Sc population in three radial regions about the original seed: $0<r<100$~nm, $100<r<175$~nm, and $175<r<250$~nm. (a) Number density, (b) number-average precipitate radius, and (c) precipitate volume fraction. The original seeded precipitate is excluded from all three quantities, while the radial-region volumes are retained as fixed geometric volumes.}
    \label{fig:heterogeneous_kinetics}
\end{figure}

The resulting microstructural heterogeneity persists after the long-wavelength composition field has largely relaxed. Figure~\ref{fig:heterogeneous_kinetics} separates the secondary population into the radial regions $0<r<100$~nm, $100<r<175$~nm, and $175<r<250$~nm, with the original seed excluded from all quantities. The two outer regions develop number densities of approximately $4.5$--$4.8\times10^{22}$~m$^{-3}$ and volume fractions of approximately 0.95\%, whereas the inner region reaches approximately $2.9\times10^{22}$~m$^{-3}$ and 0.6\%, respectively. Once precipitation is established, the mean particle radii in the three regions converge towards approximately $3.6$~nm. The persistent effect of the transient depletion field is therefore primarily a reduction in the number of secondary precipitates formed near the original seed, and consequently in the local precipitate volume fraction, rather than a substantial difference in their eventual characteristic size.

At longer times the particle number density decreases while the mean radius increases, indicating the onset of coarsening. At $t=18.04\times10^3$~s, the secondary population has $\overline{R}\simeq3.64$~nm and the mean matrix composition is approximately 0.01253~at.\% Sc. The Gibbs--Thomson equilibrium composition corresponding to this radius is 0.01264~at.\%, while the observed matrix composition corresponds to a critical radius of approximately $3.70$~nm. The late-time state is therefore consistent with curvature-controlled coarsening.

This example demonstrates the coupling between transient long-range solute redistribution and local precipitation kinetics that motivates the scale-separated formulation. Growth of the initial precipitate generates a long-wavelength depletion field that suppresses subsequent nucleation in its vicinity. Although this concentration heterogeneity eventually relaxes, the resulting differences in precipitate number density and volume fraction persist. The scale-separated formulation therefore allows an evolving mesoscale solute field to influence local precipitation kinetics without requiring that spatial heterogeneity be prescribed in advance.

\section{Conclusions}
\label{sec:conclusions}

A scale-separated monopole method has been developed for precipitation problems involving diffusion over widely separated length scales. The direct quasistatic formulation accurately reproduces single-particle growth and conventional population-level precipitation kinetics, but becomes inappropriate when spatially heterogeneous particle populations generate slowly relaxing long-wavelength concentration fields. The scale-separated formulation addresses this limitation by using an Ewald decomposition to retain rapidly relaxing particle-scale fields in a quasistatic short-range description while evolving the complementary long-range field by transient diffusion. Within the scale-separated regime, predictions become weakly dependent on the precise choice of matching length.

Application to Al--Sc demonstrates that localized precipitation can generate a transient long-range depletion field that modifies subsequent precipitation. Growth of a pre-existing particle suppresses secondary nucleation in the surrounding matrix, producing persistent spatial variations in precipitate number density and volume fraction even after the long-wavelength composition field has largely relaxed. The resulting method provides a route to spatially resolved precipitation modelling in systems where particle-scale diffusion fields and mesoscale solute heterogeneity must be treated simultaneously, without requiring the particle scale to be resolved throughout the simulation domain.

\section*{Acknowledgements}

The author gratefully acknowledges M.~Perez (MATEIS, INSA Lyon), C.~Sigli (Constellium), and F.~De~Geuser (SIMaP, Universit\'e Grenoble Alpes) for valuable discussions during the early development of this work. He also thanks the SIMaP group for their hospitality during his sabbatical stay and Universit\'e Grenoble Alpes for financial support through its Visiting Professor Program.

\section*{Code availability}

The source code implementing the methods described in this work is available at \url{https://github.com/sinclacw/monopreci}.

\bibliography{references.bib}

\end{document}


\maketitle

\tableofcontents

\newpage


\section{Multicomponent thermodynamic and growth formulation}
\label{sec:supp_multicomponent}

For each precipitate $i$ and independently diffusing solute component $\alpha$, the monopole solution provides an interfacial matrix composition $c_i^\alpha$ and monopole strength $Q_i^\alpha$. The component-specific diffusion-controlled interface velocity is

\begin{equation}
v_i^\alpha =
-\frac{D^\alpha Q_i^\alpha}
{4\pi R_i^2\left(c_{p_i}^{\alpha}-c_i^\alpha\right)}
\label{eq:supp_species_velocity}
\end{equation}

where $D^\alpha$ is the matrix diffusivity and $c_{p_i}^{\alpha}$ is the concentration of component $\alpha$ in precipitate phase $p_i$.

Local thermodynamic equilibrium determines the interfacial compositions. For a precipitate phase $p$ with stoichiometric coefficients $\nu_{p\alpha}$, the planar solubility product is

\begin{equation}
K_p(T)
=
\prod_{\alpha=1}^{S}
\left(X_{\mathrm{eq},p}^{\alpha}\right)^{\nu_{p\alpha}}
\label{eq:supp_solubility_product}
\end{equation}

where $X^\alpha$ is the matrix site fraction of component $\alpha$. Including the Gibbs--Thomson contribution for a spherical precipitate gives

\begin{equation}
\sum_{\alpha=1}^{S}\nu_{p_i\alpha}\ln X_i^\alpha
=
\ln K_{p_i}(T)
+
\left(\sum_{\alpha=1}^{S}\nu_{p_i\alpha}\right)
\frac{2\gamma_{p_i}v_{p_i}}{R_i k_{\mathrm B}T}
\label{eq:supp_GT_solubility_product}
\end{equation}

where $\gamma_{p_i}$ and $v_{p_i}$ are the matrix--precipitate interfacial energy and precipitate atomic volume, respectively. The interfacial compositions are additionally constrained by requiring a common interface velocity,

\begin{equation}
v_i^1=v_i^2=\cdots=v_i^S=\dot R_i.
\label{eq:supp_equal_velocity}
\end{equation}

Together with the component-specific monopole equations given in the main text, Eqs.~\eqref{eq:supp_species_velocity}--\eqref{eq:supp_equal_velocity} determine the interfacial compositions, monopole strengths, and growth or dissolution rate of each precipitate.

For the direct formulation, conservation of each solute component determines the uniform background matrix concentration. With $V_i=4\pi R_i^3/3$ and matrix volume $V_m=V-\sum_iV_i$,

\begin{equation}
\bar{c}^{\,\alpha}
=
\frac{
M_{\mathrm{tot}}^\alpha-\displaystyle\sum_i c_{p_i}^{\alpha}V_i
}{
V-\displaystyle\sum_iV_i
}
\label{eq:supp_mean_matrix_concentration}
\end{equation}

This relation is applied independently to each diffusing component. In the scale-separated formulation, conservation is instead enforced through conservative transfer between the particles and the evolving long-range concentration field, as described in Sec.~\ref{sec:supp_hybrid_numerics}.


\section{Classical nucleation model with spatial sensitivity}
\label{sec:supp_CNT}

For precipitate phase $p$ with stoichiometric coefficients $\nu_{p\alpha}$, the supersaturation is

\begin{equation}
\ln S_p
=
\sum_{\alpha=1}^{S}\nu_{p\alpha}\ln X^\alpha-\ln K_p(T)
\label{eq:supp_supersaturation}
\end{equation}

with nucleation permitted for $S_p>1$. Within the ideal-dilute thermodynamic description, the volumetric chemical driving force is

\begin{equation}
\Delta G_{v,p}
=
-\frac{k_{\mathrm B}T}{v_{\mathrm{fu},p}}\ln S_p,
\qquad
v_{\mathrm{fu},p}
=
\left(\sum_{\alpha=1}^{S}\nu_{p\alpha}\right)v_p
\label{eq:supp_driving_force}
\end{equation}

where $v_{\mathrm{fu},p}$ is the volume per precipitate formula unit. The critical radius and homogeneous nucleation barrier are

\begin{equation}
R_p^\ast=-\frac{2\gamma_p}{\Delta G_{v,p}},
\qquad
\Delta G_p^\ast=\frac{16\pi\gamma_p^3}{3\Delta G_{v,p}^2}
\label{eq:supp_CNT_barrier}
\end{equation}

A factor $f_p$ is applied to the barrier for heterogeneous nucleation, with $f_p=1$ for homogeneous nucleation. The Zeldovich factor and multicomponent attachment rate are

\begin{equation}
Z_p
=
\frac{v_{\mathrm{fu},p}}{2\pi {R_p^\ast}^{\,2}}
\left(\frac{\gamma_p}{k_{\mathrm B}T}\right)^{1/2}
\label{eq:supp_zeldovich}
\end{equation}

and

\begin{equation}
\beta_p^\ast
=
\frac{4\pi {R_p^\ast}^{\,2}}{a^4}
\left[
\sum_{\alpha=1}^{S}
\frac{X_p^\alpha}{D^\alpha X^\alpha}
\right]^{-1},
\qquad
X_p^\alpha
=
\frac{\nu_{p\alpha}}{\displaystyle\sum_{\beta=1}^{S}\nu_{p\beta}}
\label{eq:supp_attachment_rate}
\end{equation}

The steady-state nucleation rate is

\begin{equation}
J_p^{\mathrm{ss}}
=
N_{0,p}Z_p\beta_p^\ast
\exp\left(-\frac{f_p\Delta G_p^\ast}{k_{\mathrm B}T}\right).
\label{eq:supp_steady_nucleation_rate}
\end{equation}

For a nucleation region of available volume $V_{\mathrm{avail}}$, a reference rate $J_{p,\mathrm{ref}}$ is used to generate candidate events according to

\begin{equation}
N_{\mathrm{trial},p}
\sim
\mathrm{Poisson}\left(J_{p,\mathrm{ref}}V_{\mathrm{avail}}\Delta t\right).
\label{eq:supp_poisson_sampling}
\end{equation}

Candidate positions are sampled randomly within the region and rejected if they overlap existing precipitates. At each admissible position $\mathbf{x}$, the local matrix composition is reconstructed and the corresponding rate $J_{p,\mathrm{loc}}(\mathbf{x})$ evaluated. The candidate is accepted with probability

\begin{equation}
P_{\mathrm{acc}}(\mathbf{x})
=
\min\left[
1,\frac{J_{p,\mathrm{loc}}(\mathbf{x})}{J_{p,\mathrm{ref}}}
\right].
\label{eq:supp_local_thinning}
\end{equation}

Accepted nuclei are introduced at

\begin{equation}
R_{\mathrm{nuc},p}
=
R_p^\ast
+
\frac{1}{2}
\sqrt{\frac{k_{\mathrm B}T}{\pi\gamma_p}}
\label{eq:supp_insertion_radius}
\end{equation}

and subsequently evolve as members of the explicit precipitate population. The timestep is restricted when necessary to limit the expected number of proposed nucleation events.


\section{Transient finite-difference reference calculations}
\label{sec:supp_fd}

Reference calculations were performed using a three-dimensional sharp-interface finite-difference model based on the ShIFuMo formulation of Eymann et al. (Acta Materialia 279 (2024) 120296). These calculations provide a fully transient reference for assessing the quasistatic monopole approximation. The same Al--Sc thermodynamic and kinetic parameters were used in both descriptions.

The periodic simulation domain is discretized on a uniform Cartesian grid with spacing $\Delta x$. Each precipitate is represented by a spherical shape function with continuously evolving radius $R_i$, while grid cells are classified as matrix or precipitate according to their cell-centre positions. Solute transport in the matrix is governed by

\begin{equation}
\frac{\partial c}{\partial t}
=
D_{\mathrm{Sc}}\nabla^2c
\label{eq:supp_fd_diffusion}
\end{equation}

and is discretized using a seven-point Laplacian and backward-Euler time integration. The equilibrium matrix composition $c_i$ at each precipitate surface is determined using the same Gibbs--Thomson thermodynamics as in the monopole calculations and imposed as a Dirichlet boundary condition.

For a matrix cell adjacent to precipitate $i$, the flux through a matrix--precipitate face is evaluated as

\begin{equation}
j_f
=
D_{\mathrm{Sc}}
\frac{c_{\mathbf n}-c_i}{\Delta x/2}
\label{eq:supp_fd_faceflux}
\end{equation}

and the total flux to the particle is obtained by summing over its matrix--precipitate faces,

\begin{equation}
\mathcal J_i
=
\sum_{f\in S_i}j_f\Delta x^2.
\label{eq:supp_fd_totalflux}
\end{equation}

The continuously defined particle volume is then updated from the interfacial solute balance,

\begin{equation}
V_i^{k+1}
=
V_i^k
+
\frac{\mathcal J_i\Delta t}{c_p-c_i},
\qquad
R_i^{k+1}
=
\left(\frac{3V_i^{k+1}}{4\pi}\right)^{1/3}.
\label{eq:supp_fd_particle_update}
\end{equation}

As the radius changes, the discrete particle mask is reconstructed. The timestep is restricted such that $|\Delta R_i|\leq\eta_R\Delta x$.

Because the continuously defined spherical particle volume does not in general coincide exactly with its discrete grid representation, solute conservation is monitored using

\begin{equation}
M_{\mathrm{tot}}
=
\sum_{\mathbf n}c_{\mathbf n}\Delta x^3
+
\sum_iV_i(c_p-c_i).
\label{eq:supp_fd_inventory}
\end{equation}

Small inventory changes associated with changes in the discrete particle mask and Gibbs--Thomson interface composition are redistributed to the matrix such that $M_{\mathrm{tot}}$ remains constant to numerical precision.


\section{Finite-difference validation for interacting precipitates}
\label{sec:supp_two_particle_validation}

The direct monopole formulation was additionally compared with the transient finite-difference model for two closely spaced Al$_3$Sc precipitates with overlapping diffusion fields. Two precipitates of initial radius $R_0=2$~nm and centre-to-centre separation 8~nm were placed near the centre of a $60\times60\times60$~nm$^3$ periodic cell containing an initially homogeneous Al--0.20~at.\% Sc matrix. The system was held isothermally at $300\,^\circ$C with nucleation disabled. Calculations performed over a wider range of particle separations gave similar results.

\begin{figure*}[ht!]
\centering
\includegraphics[width=0.90\textwidth]{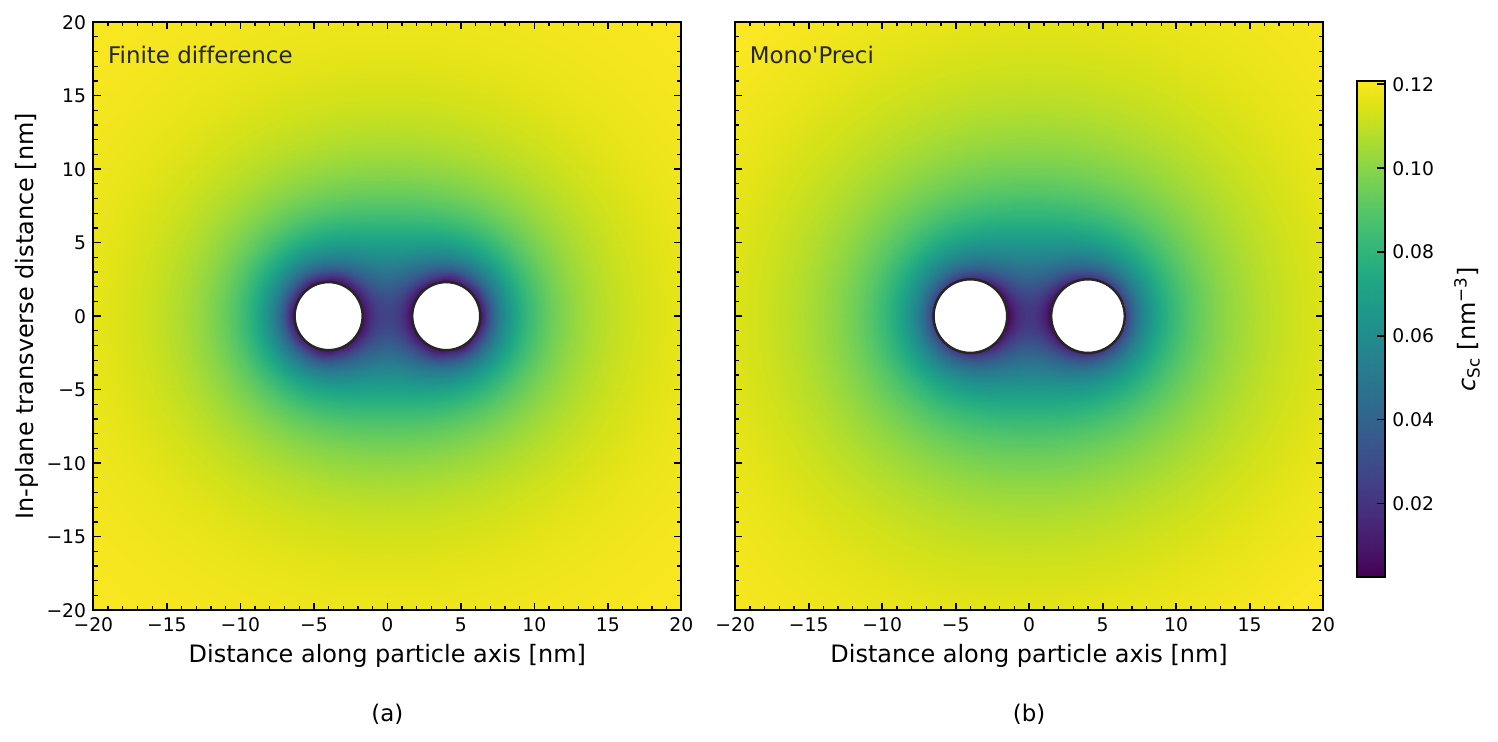}
\caption{Sc concentration field surrounding two interacting Al$_3$Sc precipitates at $t=1000$~s calculated using (a) the transient finite-difference model and (b) the direct monopole formulation. The same concentration scale is used in both panels.}
\label{fig:supp_two_particle_fields}
\end{figure*}

Figure~\ref{fig:supp_two_particle_fields} compares the Sc concentration fields at $t=1000$~s. Both calculations predict a common depleted region surrounding the growing particles, including the merging of their individual diffusion fields between the particle centres. The spatial extent and magnitude of the depletion are closely reproduced by the direct monopole calculation.

The evolution toward the quasistatic solution is quantified by comparing the two concentration fields pointwise. Defining the concentration error at sampling point $n$ as

\begin{equation}
e_n=c_{\mathrm{M},n}-c_{\mathrm{FD},n},
\label{eq:supp_pointwise_error}
\end{equation}

the total root-mean-square error is

\begin{equation}
\mathrm{RMSE}
=
\left(\frac{1}{N_g}\sum_{n=1}^{N_g}e_n^2\right)^{1/2},
\label{eq:supp_total_rmse}
\end{equation}

where $N_g$ is the number of sampled matrix points. The mean bias is

\begin{equation}
\bar e
=
\left\langle c_{\mathrm M}\right\rangle
-
\left\langle c_{\mathrm{FD}}\right\rangle
=
\frac{1}{N_g}\sum_{n=1}^{N_g}e_n,
\label{eq:supp_mean_bias}
\end{equation}

and the fluctuation root-mean-square error after removal of this mean offset is

\begin{equation}
\mathrm{RMSE}_{\mathrm{fluc}}
=
\left[
\frac{1}{N_g}
\sum_{n=1}^{N_g}(e_n-\bar e)^2
\right]^{1/2}.
\label{eq:supp_fluctuation_rmse}
\end{equation}

These quantities satisfy

\begin{equation}
\mathrm{RMSE}^2
=
\mathrm{RMSE}_{\mathrm{fluc}}^2+\bar e^2.
\label{eq:supp_rmse_decomposition}
\end{equation}

Here $|\bar e|$ measures disagreement in the spatially averaged concentration, while $\mathrm{RMSE}_{\mathrm{fluc}}$ measures disagreement in the spatial structure after removal of this mean offset.

\begin{figure*}[ht!]
\centering
\includegraphics[width=0.92\textwidth]{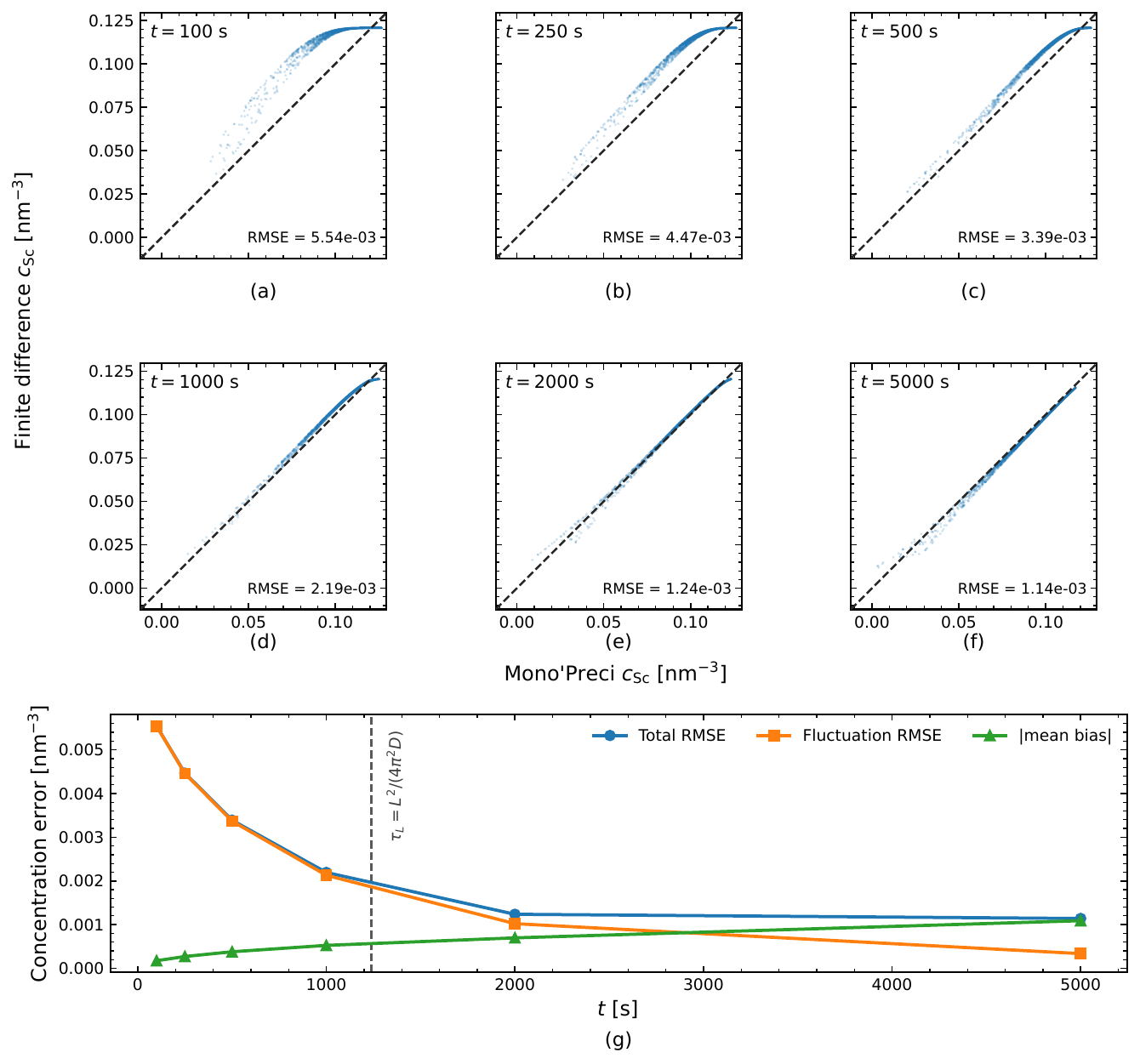}
\caption{Pointwise comparison of the Sc concentration fields calculated using the direct monopole and transient finite-difference descriptions for two interacting precipitates separated initially by 8~nm. (a--f) Finite-difference concentration plotted against the corresponding monopole concentration at $t=100$, 250, 500, 1000, 2000, and 5000~s. The dashed diagonal denotes exact agreement. (g) Total root-mean-square error $\mathrm{RMSE}$, fluctuation root-mean-square error $\mathrm{RMSE}_{\mathrm{fluc}}$, and absolute mean bias $|\bar e|$. The vertical dashed line indicates the relaxation time $\tau_L=L^2/(4\pi^2D)$ of the longest non-zero Fourier mode of the periodic cell.}
\label{fig:supp_two_particle_error}
\end{figure*}

Figure~\ref{fig:supp_two_particle_error} shows that the total RMSE decreases from $5.54\times10^{-3}$~nm$^{-3}$ at $t=100$~s to $2.19\times10^{-3}$~nm$^{-3}$ at $t=1000$~s and $1.14\times10^{-3}$~nm$^{-3}$ at $t=5000$~s. At early times the error is dominated by $\mathrm{RMSE}_{\mathrm{fluc}}$, reflecting the transient development of the depletion field in the finite-difference calculation compared with its quasistatic representation in the direct monopole formulation. This contribution decreases over times comparable to the relaxation time of the longest non-zero diffusion mode,

\begin{equation}
\tau_L=\frac{L^2}{4\pi^2D}.
\label{eq:supp_box_diffusion_time}
\end{equation}

The comparison confirms that the coupled monopole solution reproduces the overlapping concentration fields of interacting precipitates once the initial transient establishment of the diffusion field has relaxed.


\section{Numerical implementation of the scale-separated formulation}
\label{sec:supp_hybrid_numerics}

The scale-separated formulation described in the main text requires conservative transfer of solute between the discrete precipitates and the long-range Cartesian grid. For each component $\alpha$, an inventory change $\Delta M_i^\alpha$ associated with precipitate $i$ is distributed to the eight surrounding periodic grid points using trilinear interpolation weights $w_{ig}$ satisfying

\begin{equation}
\sum_g w_{ig}=1.
\label{eq:supp_weight_normalization}
\end{equation}

The corresponding unfiltered concentration increment at grid point $g$ is

\begin{equation}
\Delta c_g^\alpha
=
-\frac{1}{\Delta V}
\sum_i w_{ig}\Delta M_i^\alpha,
\label{eq:supp_grid_deposition}
\end{equation}

where $\Delta V$ is the grid-cell volume. Equation~\eqref{eq:supp_weight_normalization} ensures that the integrated change in matrix inventory is equal and opposite to the change in precipitate inventory, apart from floating-point roundoff.

The deposited field is transformed to Fourier space and multiplied by the Gaussian Ewald filter,

\begin{equation}
W_\ell(k)
=
\exp\left(-\frac{k^2\ell^2}{4}\right).
\label{eq:supp_ewald_filter}
\end{equation}

Because $W_\ell(0)=1$, filtering preserves the zero Fourier mode and hence the transferred solute inventory. The long-range concentration required at each particle position is evaluated from the grid using the corresponding trilinear interpolation.

Short-range interactions are evaluated periodically using minimum-image particle separations. Since the complementary-error-function contribution decays rapidly beyond the matching length, interactions are retained for particle pairs satisfying

\begin{equation}
r_{ij}\leq r_{\mathrm{cut}},
\end{equation}

with $r_{\mathrm{cut}}=3\ell$ for the calculations reported here. This produces a sparse short-range interaction matrix, while the spatially uniform rank-one contribution described in the main text is retained explicitly.

Long-range diffusion is performed spectrally. Particle--field coupling and diffusion are combined symmetrically over each timestep: the long-range field is first diffused for $\Delta t/2$, the particles are advanced and their inventory changes transferred to the grid, and the field is then diffused for a further $\Delta t/2$. The long-range field is retained continuously throughout the thermal history; changes in temperature modify the diffusivities and thermodynamic boundary conditions but do not reinitialize the field.


\section{Numerical sensitivity of the scale-separated formulation}
\label{sec:supp_hybrid_sensitivity}

Two numerical parameters associated with the scale-separated formulation were examined separately. Varying the matching length $\ell$ tests sensitivity to the artificial division between the quasistatic short-range and transient long-range fields, while varying the long-range grid spacing $\Delta x$ at fixed $\ell$ tests whether the transient field is adequately spatially resolved.

\begin{figure*}[ht!]
\centering
\includegraphics[width=0.90\textwidth]{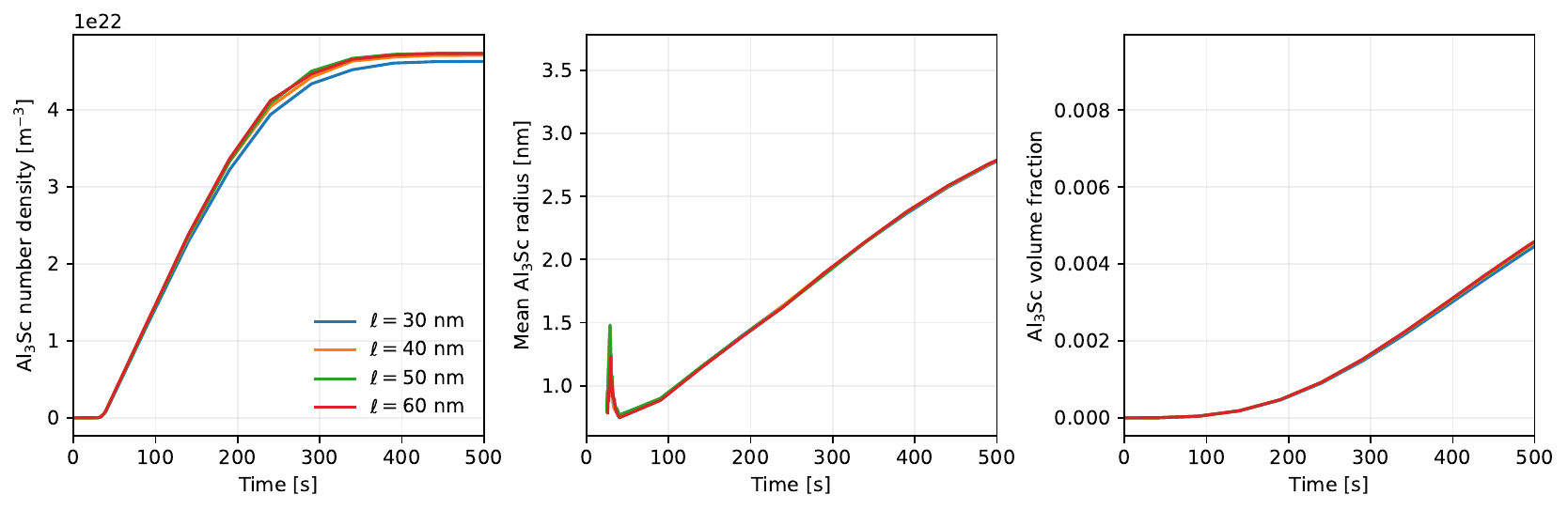}
\caption{Sensitivity of the global precipitation kinetics to the matching length $\ell$ for the heterogeneous precipitation calculation. The matching length is varied from 30 to 60~nm while all other physical and numerical parameters are unchanged.}
\label{fig:supp_ell_global}
\end{figure*}

The matching-length sensitivity was examined for the heterogeneous precipitation calculation described in the main text using $\ell=30$--60~nm while retaining the same physical and numerical conditions. Figure~\ref{fig:supp_ell_global} compares the resulting global precipitation kinetics. The predicted number density, mean precipitate radius, and volume fraction are only weakly affected by the choice of matching length. The largest difference occurs in the number density, for which the $\ell=30$~nm calculation gives a slightly lower value after nucleation has largely ceased. The mean-radius and volume-fraction histories are essentially coincident over the range examined.

\begin{figure*}[ht!]
\centering
\includegraphics[width=0.75\textwidth]{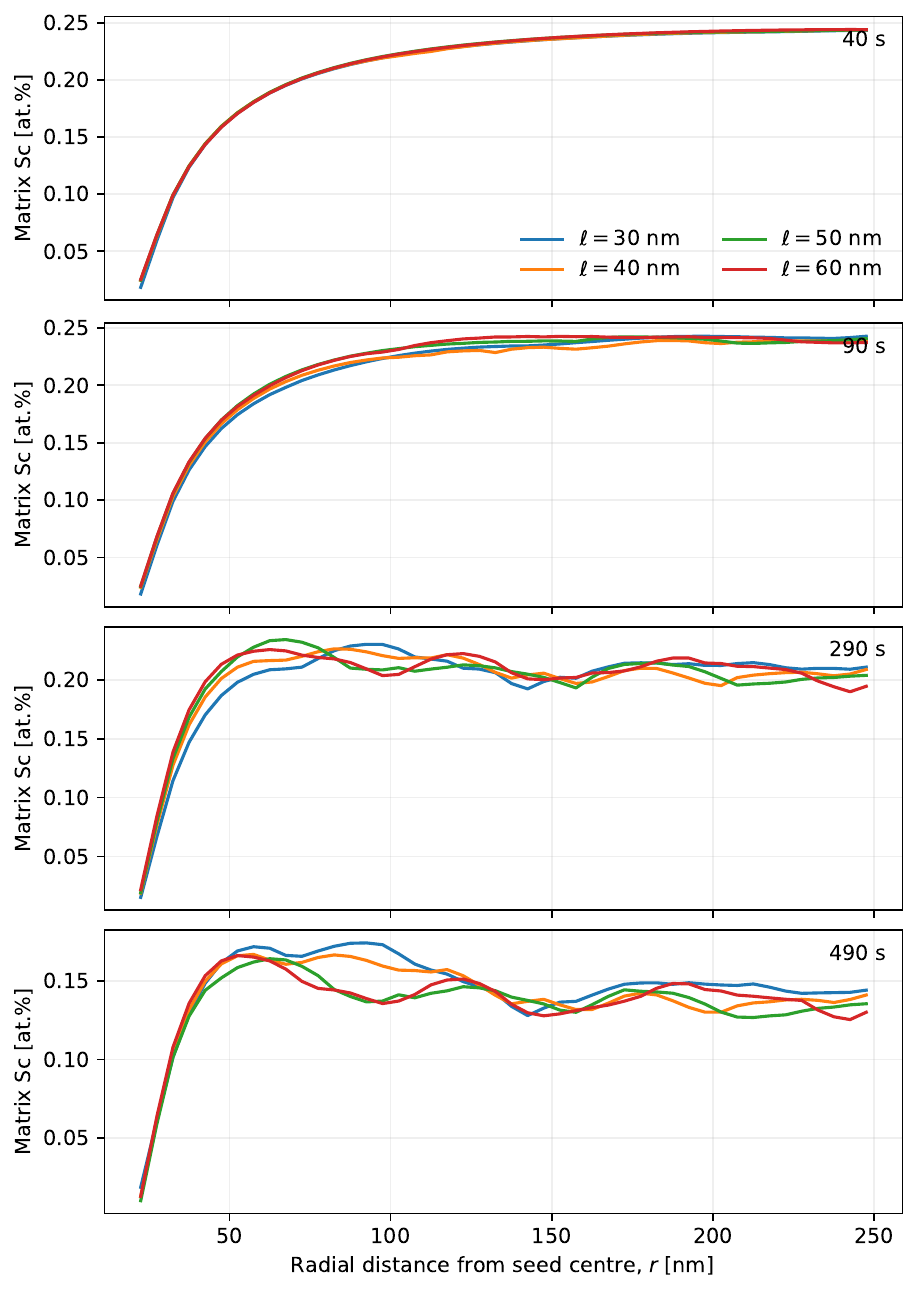}
\caption{Sensitivity of the radial matrix Sc concentration profile to the matching length $\ell$ at $t=40$, 90, 290, and 490~s. Radial distance is measured from the centre of the pre-existing seed precipitate.}
\label{fig:supp_ell_concentration}
\end{figure*}

The spatial concentration field is similarly insensitive to the matching length. Figure~\ref{fig:supp_ell_concentration} shows radial profiles of the matrix Sc concentration at four stages of the calculation. At early times the profiles are nearly indistinguishable. At later times small differences develop as the spatially heterogeneous precipitate population evolves, but the extent and magnitude of the Sc-depleted region remain similar for all four matching lengths. No systematic displacement of the depletion profile with $\ell$ is observed.

These results indicate that the predicted precipitation kinetics and spatial solute redistribution are insensitive to the precise SR/LR division over the range $\ell=30$--60~nm examined here.

The sensitivity to the spatial resolution of the long-range field was examined separately at fixed matching length, $\ell=40$~nm. The production calculation used $\Delta x=5$~nm, corresponding to $\Delta x/\ell=0.125$. A second calculation was performed with $\Delta x=10$~nm, for which $\Delta x/\ell=0.25$, while retaining the same physical parameters and simulation domain. The comparison was restricted to the first 500~s, over which both calculations were available.

\begin{figure}[t]
\centering
\includegraphics[width=0.82\linewidth]{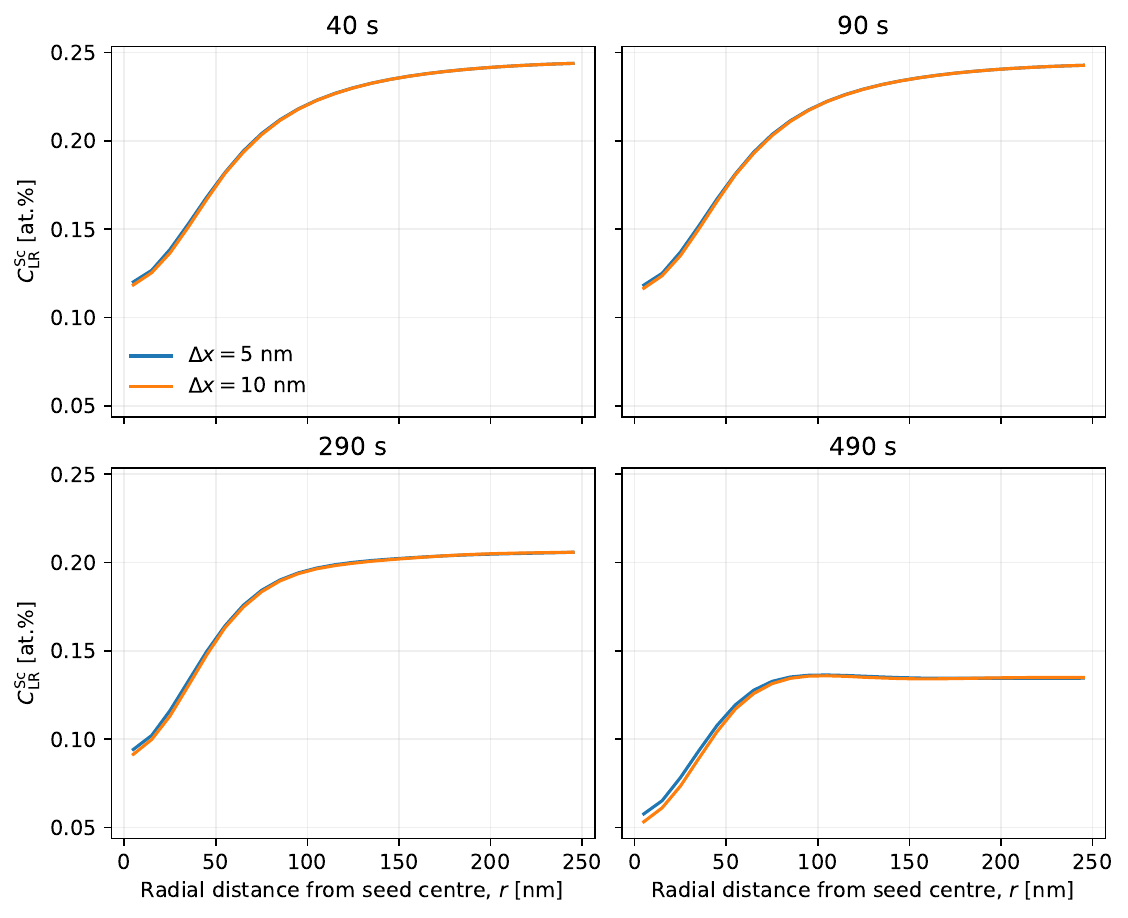}
\caption{Sensitivity of the long-range Sc concentration field to the long-range grid spacing for the heterogeneous precipitation calculation with $\ell=40$~nm. Radial profiles are spherical averages of the three-dimensional long-range field about the initial central precipitate. Results obtained with the production grid, $\Delta x=5$~nm, are compared with a calculation using $\Delta x=10$~nm.}
\label{fig:supp_dx_concentration}
\end{figure}

Figure~\ref{fig:supp_dx_concentration} compares radial profiles of the native three-dimensional long-range Sc field at four stages of the calculation. The profiles were obtained by spherical averaging about the initial central precipitate. The two grid resolutions give nearly indistinguishable profiles throughout the calculation. The root-mean-square difference between the profiles, normalized by the concentration range $\max_r C_{\mathrm{LR}}-\min_r C_{\mathrm{LR}}$ of the reference $\Delta x=5$~nm profile, is approximately $0.13\%$ at 40 and 90~s, $0.96\%$ at 290~s, and $2.17\%$ at 490~s. The corresponding maximum absolute differences remain below $2.3\times10^{-3}$~at.\% Sc.

The global precipitation kinetics are similarly insensitive to the grid spacing. The number-density, mean-radius, and volume-fraction histories obtained with $\Delta x=5$ and 10~nm are essentially coincident over the common simulation interval. Together with the matching-length sensitivity shown above, these results indicate that the production choices $\ell=40$~nm and $\Delta x=5$~nm provide a well-resolved scale separation for the heterogeneous precipitation calculation.